\documentclass[lettersize,journal]{IEEEtran}
\usepackage{amsmath,amsfonts}
\usepackage{algorithmic}
\usepackage{algorithm}
\usepackage{array}
\usepackage[caption=false,font=normalsize,labelfont=sf,textfont=sf]{subfig}
\usepackage{textcomp}
\usepackage{stfloats}
\usepackage{url}
\usepackage{verbatim}
\usepackage{graphicx}
\usepackage{booktabs}
\usepackage{xcolor}
\usepackage{cite}
\usepackage[hidelinks]{hyperref}
\usepackage{multirow}
\usepackage{textcomp}
\usepackage{amssymb}
\usepackage{tabularx}
\usepackage{placeins}
\usepackage{pifont}
\usepackage[edges]{forest}
\newcommand{\cmark}{\ding{51}}%
\newcommand{\xmark}{\ding{55}}%
\newcommand{\pmark}{\ding{108}}%

\usepackage{forest}
\usepackage{varwidth}
\usepackage{xcolor}

\newcommand{\ntnleaffont}{\fontsize{6}{6.6}\selectfont}   
\newcommand{\ntnplanefont}{\fontsize{6.6}{7.2}\selectfont}

\forestset{
  ntn base/.style={draw, rounded corners=1.6pt, align=left, font=\ntnleaffont,
    inner sep=1.8pt, s sep=0.45mm, l sep=3.2mm, line width=.35pt},
  ntn plane/.style={s sep=1.9mm, fill=#1!8, draw=#1!60!black,
    font=\ntnplanefont\bfseries,
    edge={-latex, #1!60!black, line width=.45pt},
    for descendants={draw=#1!45!black, fill=white, font=\ntnleaffont,
      edge={-latex, #1!55!black, line width=.35pt}}},
  ntn west/.style={grow=west, growth parent anchor=west,
    parent anchor=west, child anchor=east,
    edge path={\noexpand\path[\forestoption{edge}] (!u.parent anchor) -- +(-2.2mm,0) |- (.child anchor)\forestoption{edge label};}},
  ntn east/.style={grow'=east, growth parent anchor=east,
    parent anchor=east, child anchor=west,
    edge path={\noexpand\path[\forestoption{edge}] (!u.parent anchor) -- +(2.2mm,0) |- (.child anchor)\forestoption{edge label};}},
}

\begin{document}

\title{Semantic Communication for Non-Terrestrial Networks: A Survey from Platform Constraints to System Design}

\author{Loc X. Nguyen, Avi Deb Raha, Huy Q. Le,  Zhu Han~\IEEEmembership{Fellow,~IEEE,}\\ Eui-Nam Huh~\IEEEmembership{Senior Member,~IEEE}, and Choong Seon Hong~\IEEEmembership{Fellow,~IEEE}

\thanks{Loc X. Nguyen, Avi Deb Raha, Huy Q. Le, Eui-Nam Huh, and Choong Seon Hong are with the Department of Computer Science and Engineering, Kyung Hee University,  Yongin-si, Gyeonggi-do 17104, Rep. of Korea, e-mails:{\{xuanloc088, avi, quanghuy69, johnhuh, cshong\}@khu.ac.kr}.}

\thanks{Zhu Han is with the Department of Electrical and Computer Engineering at the University of Houston, Houston, TX 77004 USA, and also with the Department of Computer Science and Engineering, Kyung Hee University, Seoul, South Korea, 446-701. email:{\{hanzhu22\}}@gmail.com.}

}

\markboth{Journal of \LaTeX\ Class Files,~Vol.~14, No.~8, August~2021}%
{Author \MakeLowercase{\textit{et al.}}: Semantic Communication in Non-Terrestrial Networks}

\maketitle

\begin{abstract}
Sixth-generation networks are expected to extend connectivity beyond terrestrial infrastructure through non-terrestrial networks (NTNs) comprising satellites, high-altitude platform stations, and unmanned aerial vehicles. However, these platforms operate in a regime that bit-fidelity-centric design handles poorly: high free-space path loss, massive round-trip delays, Doppler shifts of hundreds of kilohertz, limited visibility windows, and limited on-board computing capability compared with ground hardware. Semantic communication (SemCom), which transmits task-relevant meaning rather than exact bits, provides a promising way to address these constraints. This survey examines SemCom for NTNs from the perspective of how semantic mechanisms support different parts of the communication system. We first map five structural NTN constraints onto the semantic mechanisms that can address them, and we show that each platform imposes a distinct constraint vector that selects among those mechanisms. We then propose a five-plane taxonomy covering semantic representation and on-board encoding, channel-adaptive transmission, semantic networking, resource management, and distributed learning with knowledge-base maintenance, together with a cross-cutting trust plane, and we review the literature within it. Finally, we summarize current standardization efforts and available research resources, and identify open problems and future research directions for SemCom in NTNs.
\end{abstract}

\begin{IEEEkeywords}
Deep joint source-channel coding, generative AI, low Earth orbit satellites, non-terrestrial networks, semantic communication, space-air-ground integrated networks.
\end{IEEEkeywords}

\vspace{-0.1in}
\section{Introduction}
\label{sec:Introduction}
Connected devices, immersive services, and latency-sensitive applications continue to push terrestrial wireless infrastructure toward its operational limits. In addition, terrestrial deployment remains economically impractical across large parts of the world. Roughly half the planet's surface will never support the capital cost of a terrestrial base station~\cite{6GSurvey,SatelliteSurveyIntro}. Non-terrestrial networks (NTNs) close this gap by lifting the access node off the ground. A modern NTN is a layered system of low Earth orbit (LEO), medium Earth orbit (MEO), and geostationary Earth orbit (GEO) satellites, high-altitude platform stations (HAPS), unmanned aerial vehicles (UAVs), and, increasingly, maritime and deep-space nodes~\cite{MahboobSatelliteSurvey,11051254,10103768,9861699}. This is no longer a speculative trajectory. Mega-constellations already operate thousands of satellites with optical inter-satellite links (ISLs)~\cite{StarlinkConstellation}, and 3GPP has standardized NTN access from Release~17 onward~\cite{HosseinianSatelliteReview,11186253}. What has not changed is the underlying physics. NTN links can experience more than $200$~dB of free-space path loss at Ka-band GEO, one-way delays up to minutes, Doppler shifts of hundreds of kilohertz at LEO velocities~\cite{NTNTutorial}, and contact windows of only $5$-$15$~minutes. On-board computation and power are further limited by size, weight, power, and cost (SWaP-C) constraints~\cite{HosseinianSatelliteReview,10579820}. These conditions make it difficult to rely on transmitting and recovering every bit of the source data. A modern Earth-observation (EO) satellite may generate terabytes of data per day and have a few minutes of downlink in which to send them.

Semantic communication (SemCom) changes the communication goal. Following Weaver's three levels~\cite{Weaver}, it moves the focus from the technical level, where the goal is to reconstruct bits at the receiver, to the semantic and effectiveness levels, where the intended meaning of the message is conveyed and delivered compactly to the receiver~\cite {Weaver,Shannon,Niu2022Paradigm}. For example, rather than transmitting every pixel of a flood image, SemCom only sends a segmentation mask, a caption, or a set of polygon coordinates, which cuts the payload by orders of magnitude while preserving what a rescue coordinator needs. Three approaches to SemCom have emerged, which differ in what representation is placed on the channel and in how the receiver recovers meaning from it. Deep joint source-channel coding (D-JSCC) learns a direct mapping from the source to channel symbols~\cite{Bourtsoulatze2019DeepJSCC,DeepSC}. Generative-artificial-intelligence (GAI) SemCom sends a compact conditioning signal and reconstructs the output with a generative model at the receiver~\cite{GenerativeAISem,FMSat}. Theory-of-mind (ToM) and knowledge-graph SemCom exchange structured content that the receiver completes by reasoning over a shared knowledge base (KB)~\cite{ChristinaLessdata,ThomasReason1}.

\vspace{-0.1in}
\subsection{Motivation}
\label{subsec:motivation}

SemCom is particularly relevant to NTNs because several of its core mechanisms address the constraints that distinguish non-terrestrial communication. Specifically, task-oriented SemCom compression makes better use of a short contact window, since it sends only task-relevant features. Moreover, D-JSCC avoids the cliff effect that separate source and channel coding suffers once a Ka-band link falls below its design point. Generative decoding rebuilds high-quality data from a small conditioning signal that can survive a rain fade. Asymmetric encoder-decoder designs put the heavy computation on the ground rather than on the platform. Terrestrial networks normally assume a stable channel, comparable computing at both ends, and a single hop. However, these assumptions do not hold in NTNs, which leads to three new problems: channel adaptation must account for predictable orbital motion, semantic representations may need to traverse multiple hops without repeated re-encoding, and the network must operate around intermittent connectivity and changing network topology. Therefore, this survey examines how SemCom must be designed differently under NTNs.

\begin{table*}[t]
\centering
\caption{Comparison with existing surveys on SemCom and NTNs. \cmark: covered in depth; \pmark: partially covered; \xmark: not covered.}
\label{tab:RelatedSurvey}
\scriptsize
\renewcommand{\arraystretch}{1.15}
\setlength{\tabcolsep}{3pt}
\begin{tabular}{|>{\centering\arraybackslash}p{1.15cm}|>{\centering\arraybackslash}p{1.55cm}|c|c|c|c|c|c|c|p{9.8cm}|}
\hline
\textbf{Ref., Year} & \textbf{NTN platforms} & \textbf{REP} & \textbf{CH} & \textbf{NET} & \textbf{RES} & \textbf{LRN} & \textbf{SEC} & \textbf{STD} & \multicolumn{1}{c|}{\textbf{Scope and principal limitation with respect to NTN-SemCom}} \\ \hline

\cite{SemanticSurvey4}, 2023 & None & \cmark & \pmark & \xmark & \pmark & \pmark & \pmark & \xmark & Reviews SemCom fundamentals and applications. Entirely terrestrial; no orbital dynamics. \\ \hline

\cite{ZLuSurvey}, 2024 & None & \cmark & \pmark & \pmark & \pmark & \pmark & \xmark & \pmark & Establishes why significance and effectiveness matter. Foundational rather than platform-specific. \\ \hline

\cite{NTNTutorial}, 2025 & LEO/GEO/ HAPS/UAV & \xmark & \xmark & \pmark & \cmark & \xmark & \pmark & \cmark & NTN tutorial covering edge computing, trajectories, and radio access. Semantics absent from its design space. \\ \hline

\cite{10485510}, 2024 & LEO/UAV & \pmark & \pmark & \xmark & \pmark & \xmark & \pmark & \xmark & Magazine-length vision naming multi-modality, security, and allocation as open problems. Positional, not a review. \\ \hline

\cite{SemanticSurvey0}, 2025 & UAV only & \cmark & \pmark & \xmark & \pmark & \pmark & \cmark & \xmark & SemCom architecture, security, and privacy. NTN appears only as a UAV example. \\ \hline

\cite{10670196}, 2024 & SAGIN & \pmark & \xmark & \pmark & \cmark & \xmark & \cmark & \xmark & Generative AI across SAGIN, with SemCom one application among several. Thin on D-JSCC and knowledge graphs. \\ \hline

\cite{SemanticSurvey6}, 2023 & None & \pmark & \pmark & \xmark & \pmark & \xmark & \xmark & \xmark & Metrics for semantic and goal-oriented communication and no platform analysis. \\ \hline

\cite{ChristinaLessdata}, 2025 & None & \cmark & \pmark & \pmark & \pmark & \cmark & \pmark & \pmark & Reasoning-centric formulation that explicitly sets aside D-JSCC and generative decoding. NTN not considered. \\ \hline

\cite{MengSurvey}, 2025 & Space-air-ground-sea & \cmark & \pmark & \pmark & \pmark & \pmark & \pmark & \pmark & First survey of SemCom for SAGSIN, organized by significance, meaning, and effectiveness. Thorough on D-JSCC, thinner elsewhere. \\ \hline

\cite{Hu2025SAGIN}, 2026 & SAGIN & \pmark & \xmark & \pmark & \pmark & \pmark & \cmark & \pmark & Generative-AI-empowered secure SAGIN communication, organized by threat class. Security-first rather than design-plane. \\ \hline

\cite{Ahmed2025MultiSemCom}, 2025 & Satellite & \cmark & \pmark & \xmark & \cmark & \xmark & \pmark & \pmark & Semantic and bit-level coexistence through NOMA. Narrow in plane coverage by design. \\ \hline

\cite{SatelliteSurveyIntro}, 2026 & Satellite/UAV & \pmark & \xmark & \pmark & \cmark & \pmark & \cmark & \cmark & Agentic and generative AI for satellite-augmented networks, with a strong resource inventory. Semantic transmission peripheral. \\ \hline

\cite{ZhangSemRA}, 2026 & Mostly terrestrial & \pmark & \pmark & \xmark & \cmark & \pmark & \pmark & \xmark & Resource allocation in wireless SemCom from framework to solution families. Terrestrial throughout. \\ \hline

\cite{FontanesiAI}, 2025 & GEO/MEO/ LEO/HAPS & \xmark & \pmark & \pmark & \cmark & \pmark & \pmark & \cmark & AI across satellite use cases including beamforming, routing, and security. Semantics not included \\ \hline

\cite{EnergyAwareSemComSat}, 2026 & LEO & \pmark & \pmark & \xmark & \cmark & \xmark & \xmark & \xmark & Energy-efficient SemCom for LEO satellites only. Single plane, single tier. \\ \hline

\textbf{This survey} & \textbf{GEO/MEO/ HAPS/LEO /UAV/deep space} & \cmark & \cmark & \cmark & \cmark & \cmark & \cmark & \cmark & \textbf{Organizes the NTN-SemCom literature by the design plane at which the semantic mechanism intervenes so that works sharing a problem structure are compared directly. Maps five constraints onto the semantic mechanisms that address them, derives platform constraint vectors that explain design choices, and consolidates standardization status.} \\ \hline
\end{tabular}
\end{table*}

\subsection{Related Surveys and the Gap This Survey Fills}
\label{subsec:related}

Existing surveys have studied SemCom and NTN extensively, but their intersection remains less explored. The SemCom literature is now mature, ranging from broad architectural surveys~\cite{SemanticSurvey4,WangSurvey,GetuFullSemanticSurvey} and discussions of what SemCom should achieve~\cite{ZLuSurvey,ChristinaLessdata} to focused reviews of metrics~\cite{SemanticSurvey6}, resource allocation~\cite{ZhangSemRA}, reinforcement learning~\cite{yan2025review}, and security~\cite{SemanticSurvey0,DWonSurvey,SecureSemSurvey}. NTN and satellite networking have developed in parallel, with surveys covering the evolution from 5G to 6G NTN~\cite{9861699,HosseinianSatelliteReview}, tutorials~\cite{NTNTutorial}, physical-layer techniques~\cite{10542348,10179219}, and AI for satellite networks~\cite{FontanesiAI,MahboobSatelliteSurvey}. These works examine topics such as hardware, MIMO processing, routing, and resource optimization in depth, but generally do not consider semantics as part of the network design.

Only a few surveys directly address NTN-SemCom, and each covers a different part of the field. Peng \emph{et al.}~\cite{10485510} provided an early architectural vision, but the work is mainly positional rather than a systematic review of the literature. Meng \emph{et al.}~\cite{MengSurvey} presented a substantial survey of SemCom in space-air-ground-sea integrated networks, organized around significance, meaning, and effectiveness. It provides detailed coverage of D-JSCC, but gives less attention to generative and knowledge-graph approaches and NTN-specific networking. Zhang \emph{et al.}~\cite{10670196} and Hu \emph{et al.}~\cite{Hu2025SAGIN} focus on generative AI for SAGIN, with the latter emphasizing security; in both cases, semantic transmission is one application among several. Other reviews have narrower scopes, such as semantic-bit coexistence over satellite links~\cite{Ahmed2025MultiSemCom} and energy-efficient SemCom for LEO satellites~\cite{EnergyAwareSemComSat}.

These surveys leave two main gaps. The first is \textbf{coverage}: no existing survey covers all five design planes across different NTN platforms, and some focus on only a single SemCom paradigm. The second, and more important, gap is \textbf{organization}. Existing NTN-SemCom surveys are often organized by platform, which separates works that address the same underlying problem. For example, the trade-off between compression and computation energy remains the same whether semantic encoding is performed on a satellite, HAPS, or UAV. Existing surveys provide relatively strong coverage of resource management and security, but give less attention to semantic networking, distributed learning, and knowledge-base maintenance. Table~\ref{tab:RelatedSurvey} summarizes these gaps. In the table, REP-SEC corresponds to the five design planes introduced in Section~\ref{subsec:taxonomy} and the cross-cutting trust plane: REP denotes semantic representation and on-board encoding; CH, channel-adaptive transmission; NET, semantic networking and multi-hop transport; RES, resource management and cross-layer optimization; LRN, distributed learning and knowledge-base maintenance; and SEC, security, privacy, and safety. STD indicates coverage of standardization and open resources.

\begin{figure*}[t]
\centering
\includegraphics[width=0.9\textwidth]{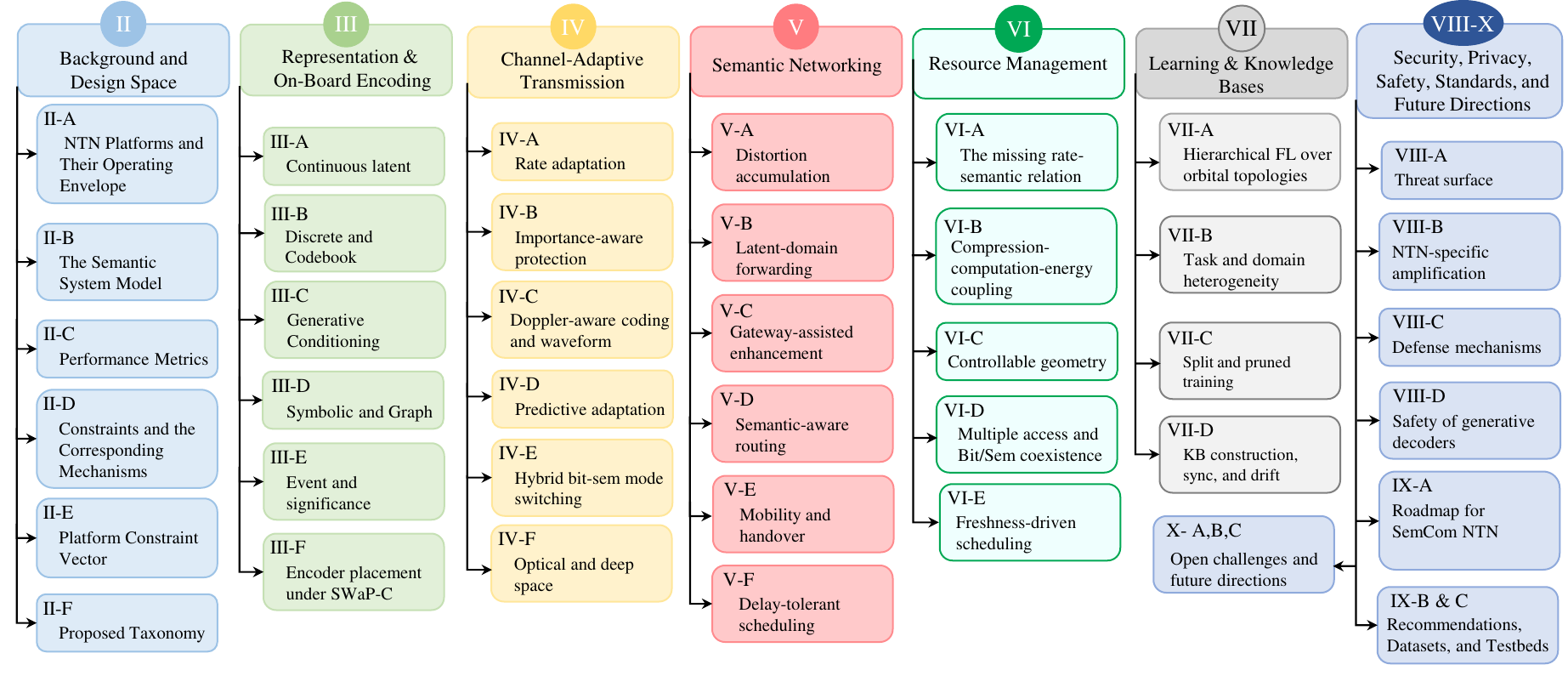}%
\caption{Roadmap of this survey.}
\label{fig:Structure}
\vspace{-0.1in}
\end{figure*}

\subsection{Contributions}
\label{subsec:contributions}

The main contributions of this survey can be illustrated as follows:
\begin{itemize}
\item We identify five structural constraints in NTNs: adverse and time-varying propagation, delay and limited contact time, SWaP-C limits, multi-hop heterogeneity, and the absence of a single trusted operator. For each constraint, we examine the semantic mechanisms used to address it and the costs they introduce. We then show that these constraints affect NTN platforms differently, forming a distinct \emph{constraint vector} for each platform. This vector, rather than the platform type alone, helps determine which semantic mechanisms are most suitable.

\item We organize the literature according to the design plane in which each semantic mechanism operates. Unlike platform-based classifications, this taxonomy places works that address the same design problem in the same section. At the same time, we retain the NTN platform as a separate dimension in every summary table, allowing the literature to be examined from either perspective.

\item Within each design plane, we first define the classification criterion and then organize the literature according to it. We connect the reviewed works through the problems they address and the limitations that remain. The summary tables capture the main dimensions that distinguish these works, including the transmitted representation, adaptation signal, operations performed by intermediate nodes, and the surrogate used to relate resource allocation to semantic quality, alongside the main limitation of each work.

\item We consolidate the 3GPP NTN roadmap and identify three hooks needed to support native semantic capabilities. We also review the datasets, simulators, and testbeds available for reproducible NTN-SemCom research. Based on the findings from these analyses and the preceding sections, we identify five research directions, each motivated by a specific gap found in the existing literature.

\end{itemize}

\subsection{Organization}
\label{subsec:organization}

Fig.~\ref{fig:Structure} shows the structure of the survey. Section~\ref{sec:Background} covers the fundamentals of NTN and SemCom, maps the five structural NTN constraints onto the semantic mechanisms that answer them, and states the taxonomy. Sections~\ref{sec:Representation} through~\ref{sec:Security} review the five design planes and the trust plane. Section~\ref{sec:Standards} assesses standardization and open resources, Section~\ref{sec:Challenges} sets out five open challenges, and Section~\ref{sec:Conclusion} concludes.

\section{Background and Design Space}
\label{sec:Background}

This section introduces the NTN environment and the semantic communication models considered in this survey. It then identifies five structural constraints that distinguish NTNs, examines the semantic mechanisms used to address them and their associated costs, and shows how differences in the severity of these constraints across platforms lead to different design choices. Based on this analysis, we introduce the taxonomy used to organize the remainder of the survey.

\begin{figure*}[t]
\centering
\includegraphics[width=0.9\textwidth]{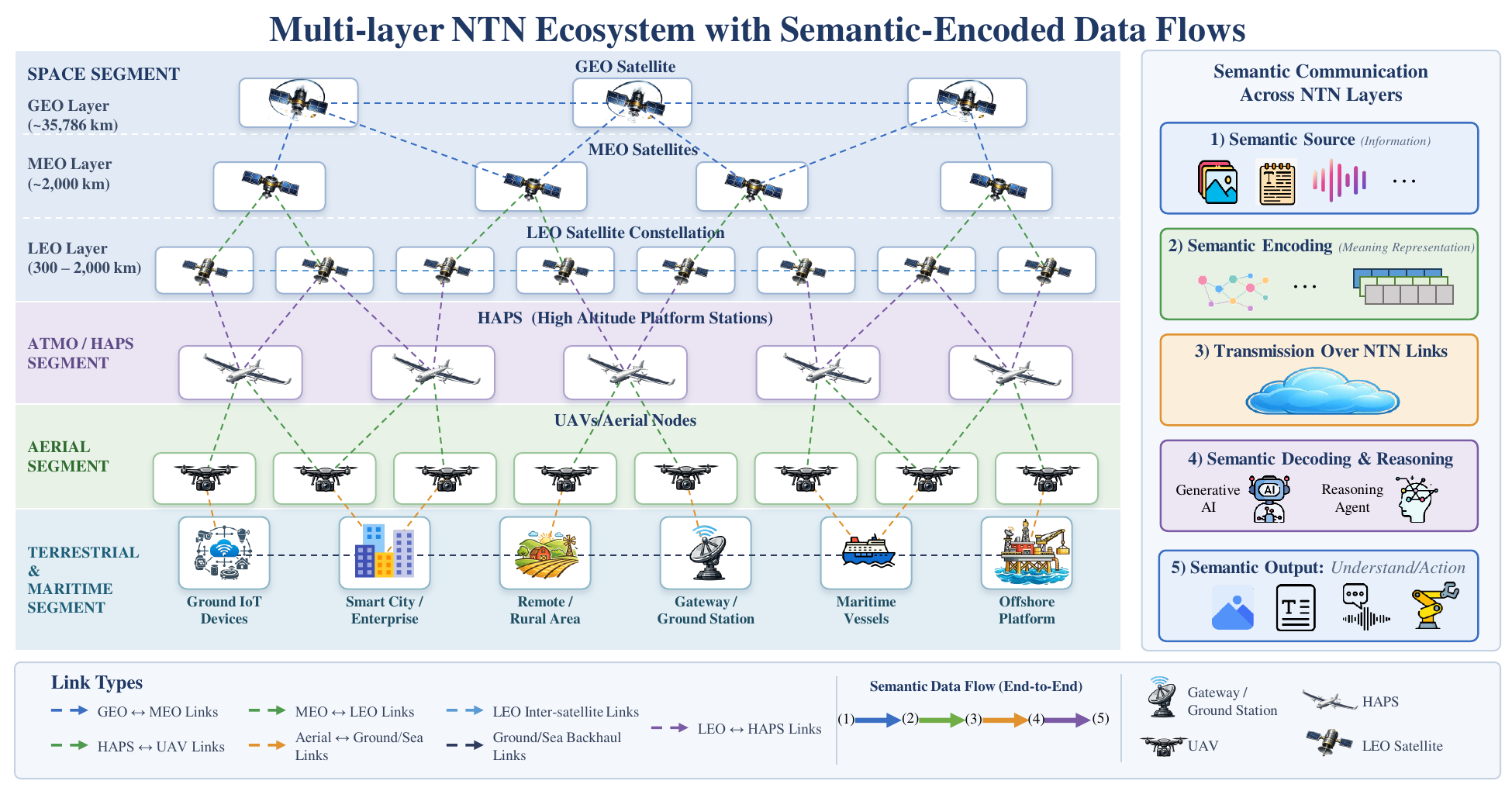}%
\caption{High-level architecture of a non-terrestrial network with semantic communication. Each tier, whether LEO, MEO, GEO, HAPS, UAV, or ground and sea, acts at once as a sensing source and as a node in a multi-hop semantic relay chain. The same semantic latent can traverse radio access links, optical inter-satellite links, and feeder links with only the channel codec changing at each hop, which is the property that makes semantic representations attractive as a common currency across heterogeneous NTN segments.}
\label{fig:NTN_Overview}
\vspace{-0.1in}
\end{figure*}

\subsection{NTN Platforms and Their Operating Envelope}
\label{subsec:NTNArch}

NTNs include three main satellite orbits: LEO, MEO, and GEO, each with different coverage, delay, and mobility characteristics. LEO satellites operate at altitudes of roughly $300$-$2{,}000$~km, MEO satellites at $2{,}000$-$35{,}786$~km, and GEO satellites at approximately $35{,}786$~km~\cite{maral2020satellite,mnm_satellite_networks}. GEO satellites provide continuous coverage over a fixed area but have the longest propagation delay. In contrast, a single LEO satellite is visible for only a short period. Mega-constellations such as Starlink address this limitation by operating large numbers of LEO satellites, turning limited contact time into frequent handovers~\cite{10579728,10494421,10778303}. At lower altitudes, HAPS maintain quasi-stationary positions at $17$--$25$~km and can operate for several months while carrying larger payloads than UAVs~\cite{9380673,5995281,airbus_zephyr}. 

UAVs operate closer to the ground and differ from the other platforms because their positions can be actively controlled, although movement directly consumes flight energy~\cite{10004947}. Together, these platforms form a space-air-ground integrated network (SAGIN), or a space-air-ground-sea integrated network (SAGSIN) when maritime nodes are included~\cite{9628162,MengSurvey}, as illustrated in Fig.~\ref{fig:NTN_Overview}.

Several features of the NTN architecture directly affect where semantic processing can be performed. The division of NTN into space, ground, control, and user segments~\cite{7572177,11316666,9420293,pasandi2024survey} provides several possible processing locations, from the platform itself to ground infrastructure. Choosing where to process semantic information therefore involves a trade-off among computation, energy consumption, and communication load. The shift from \emph{transparent} bent-pipe payloads to \emph{regenerative} payloads with digital channelization, on-board beamforming, and software-defined modems~\cite{ortiz2023onboard} further increases the possibility of performing semantic processing on board. However, its practical use remains limited by the platform's available power, battery capacity, and ability to dissipate heat~\cite{8786872}. At the same time, 3GPP introduced NTN access in Releases~17 and~18~\cite{3gpp_ts23501_rel17,3gpp_ts38300_rel18}, while AI/ML-native features are expected to emerge in Release~20~\cite{11186253}. These developments raise a broader question: should SemCom remain an application built on top of the NTN architecture, or should semantic capabilities become part of the network itself?

At the physical and access layers, high-throughput satellites use digital beamforming and beam hopping to generate dozens or even hundreds of spot beams~\cite{11433662,kouki2025phased,wang2023efficient}. Interference between neighboring beams is managed through frequency and polarization reuse and multi-user precoding~\cite{10719632,7811843}. Flexible payloads provide further control by allowing the power, bandwidth, pointing direction, and hopping pattern of each beam to be adjusted through software~\cite{9758046,8976431,9469916,9968247}. Beyond conventional orthogonal multiple access, satellite systems also support space-division~\cite{6363801}, non-orthogonal~\cite{9241448,10508590,9839554}, and rate-splitting~\cite{9205852,10312769,10299599} schemes. These mechanisms provide additional flexibility for semantic-aware resource allocation~\cite{ZhangSemRA}. At the network layer, LEO mega-constellations form multi-hop networks through ISLs, which are increasingly optical and can support multi-gigabit transmission~\cite{10184196,HU2026114969,khalid2024characterization}. These links allow information to be forwarded and processed across satellites and create the possibility of carrying semantic representations over multiple hops without repeated re-encoding. However, routing becomes more difficult as the constellation grows and the network topology changes rapidly~\cite{8761611,9149175,634801}. This has motivated learning-based routing methods that can adapt to changing network conditions~\cite{8451957,li2018diffusion,9448341,11048425}. However, most existing methods still optimize conventional network metrics rather than the requirements of semantic traffic, which we provide a depth discussion in Section~\ref{sec:Networking}.

\subsection{The Semantic System Model}
\label{subsec:levels}

Weaver divided the communication problem into three levels: \emph{Level~A}, how accurately symbols are transmitted; \emph{Level~B}, how precisely the transmitted symbols convey the intended meaning; and \emph{Level~C}, how effectively the received meaning influences the desired action. Current satellite communication remains at Level~A by delivering bits as reliably as possible. However, this approach can require substantial communication resources and suffers from the cliff effect when the channel quality falls below a threshold. Therefore, Level~B systems are gaining attention and are envisioned as the next pillar of NTN communication. An early foundation was provided by Carnap and Bar-Hillel in 1952~\cite{carnap1952outline}, who developed a theory of semantic information based on logical probability. Its practical use, however, was limited by the technology available at the time. Semantic information also remains difficult to formalize because it has no universally accepted unit~\cite{11454444}, depends on the source, the receiver's task, and its prior knowledge, and requires sufficient computation to extract and interpret meaning. Deep learning-based SemCom takes a more practical approach: semantic encoders and decoders are first learned from data, and suitable metrics are then used to measure the information they preserve or the task they support~\cite{10855638,10634888}.

A key idea in SemCom is that the transmitter and receiver can use shared knowledge, typically maintained through a knowledge base (KB)~\cite{10318078}. The KB may contain prior information such as geographic knowledge, common-sense facts, vocabularies, or foundation-model parameters, together with information about the downstream task. With this shared knowledge, the transmitter can avoid sending information that the receiver can already infer. In the canonical architecture~\cite{SemanticSurvey4}, the transmitter first performs semantic encoding using the KB and then applies channel encoding. The receiver reverses this process, using semantic decoding to recover the intended meaning instead of classical source decoding. The resulting communication chain can be written as follows:
\begin{equation}
\hat{\mathbf{s}} = f_{\text{dec}}\!\big(\,h(f_{\text{enc}}(\mathbf{s};\boldsymbol{\theta}_e)) + \mathbf{n};\,\boldsymbol{\theta}_d\,\big),
\end{equation}
trained end-to-end under the power constraint $\frac{1}{k}\mathbb{E}[\Vert x\Vert^{2}]\leq P$. For reconstruction-oriented designs, the training objective is:
\begin{equation}\label{lossfunction}
\min_{\boldsymbol{\theta}_e,\boldsymbol{\theta}_d}\;\mathbb{E}_{s,h,n}\!\left[d_{\text{sem}}(s,\hat{s})\right] + \lambda\, R(\boldsymbol{\theta}_e),
\end{equation}
where $d_{\text{sem}}$ measures semantic distortion. It can be pixel-wise mean squared error, a perceptual loss, or embedding cosine distance for text, and $R$ regularizes bandwidth, latency, or model size. For goal-oriented designs, the objective is to maximize task utility as in the following equation:
\begin{equation}\label{loss}
\max_{\boldsymbol{\theta}_e,\boldsymbol{\theta}_d}\;\mathbb{E}_{s,h,n}\!\left[U\!\left(\mathcal{T}(\hat{s}),\,y_{\text{task}}\right)\right] - \mu\,\mathbb{E}\!\left[C(\boldsymbol{\theta}_e,\boldsymbol{\theta}_d)\right],
\end{equation}
where $\mathcal{T}(\cdot)$ maps the reconstruction to a task output, $U$ is a task utility such as accuracy or mission success, and $C$ is computation cost. It is worth noting that this last term is optional in terrestrial formulations but required in NTNs. We distinguish three SemCom paradigms based on what is transmitted over the channel and how the receiver recovers the source meaning~\cite{Nguyen2025Survey}. 

\emph{D-JSCC} jointly performs source compression and channel coding using a single neural network. It maps the source directly to channel symbols as $x=f_\theta(s):\mathbb{R}^{d}\!\to\!\mathbb{C}^{k}$, with bandwidth compression ratio $\rho=k/d$, under Eq.~\eqref{lossfunction}~\cite{Bourtsoulatze2019DeepJSCC}. Its main advantage is that reconstruction quality degrades smoothly as the SNR decreases, avoiding the cliff effect of conventional digital communication. A digital variant~\cite{D2JSCC} further introduces vector quantization, allowing the learned representation to be transmitted over standardized digital waveforms.

\emph{GAI SemCom} instead uses a generative model $G_\phi$ as the decoder. The transmitter sends a compact prompt or latent representation $p$, where $|p| \ll |s|$, and the receiver generates $\hat{s}\sim G_\phi(\cdot\mid \hat{p},\xi)$ rather than reconstructing the source through a deterministic inverse~\cite{rombach2022high}. Because the generative model provides strong prior knowledge, only a small amount of information needs to be transmitted. This reduction, however, comes with two costs: generative decoding can require substantial computation, and the generated output may appear plausible without being fully faithful to the original source.

\emph{ToM and knowledge-graph SemCom} take a different approach by representing the source as a semantic graph of (head, relation, tail) triplets. The transmitter sends only the information that is not already available in the receiver's knowledge base, i.e., $\Delta G = \{\phi\in G:\phi\notin\mathcal{K}\}$. The receiver then uses the shared knowledge base $\mathcal{K}$ to infer missing or corrupted triplets~\cite{KGSemUAV,ChristinaLessdata}. This approach has two useful properties for NTN. First, knowledge-base synchronization can be performed in batches and its cost spread across multiple short contact windows. Second, when an entity is corrupted during transmission, the receiver may still recover its meaning from its relations with other entities.

These paradigms differ in how information is represented and recovered. In NTNs, their applications have mainly focused on three source modalities: image, text, and speech, although their adoption differs from the broader SemCom literature. \emph{Image} SemCom is particularly common in NTNs because it naturally supports Earth observation applications~\cite{FMSat,IRST,10901032}. \emph{Text} is more widely studied in the general SemCom literature, where early recurrent architectures~\cite{8461983,8445924} have evolved into transformer-based designs such as DeepSC~\cite{DeepSC}. These methods remain robust at low SNR and can support NTN applications such as telemetry and situation reporting. In contrast, \emph{speech} SemCom, including methods based on self-supervised models such as wav2vec~2.0 and HuBERT~\cite{9450827,baevski2020wav2vec,9585401,tong2021federated}, has received relatively little attention in NTNs. This gap is important because voice communication is needed in maritime, polar, aviation, and disaster-response scenarios, where terrestrial coverage may be unavailable. Multimodal SemCom is also beginning to appear in NTNs, particularly for broadcast services~\cite{LiuFangyu}.

\subsection{Performance Metrics}
\label{subsec:metrics}
Unlike bit-level communication, SemCom has no single universal metric because the appropriate measure depends on what the system is designed to preserve or achieve. \emph{Fidelity} metrics evaluate pixel-level reconstruction using peak SNR and MS-SSIM~\cite{Bourtsoulatze2019DeepJSCC,wang2003multiscale}, perceptual quality using LPIPS and Fr\'echet inception distance~\cite{10001359}, and text similarity using BLEU and sentence-embedding distances~\cite{DeepSC}. \emph{Task} metrics instead measure the performance of the downstream application, such as classification accuracy, detection mean average precision, or segmentation intersection-over-union~\cite{VisualEventLEO,KGSemUAV}. \emph{Cost} metrics capture the resources required for transmission, including the bandwidth compression ratio, channel symbols per source symbol, and end-to-end delay. They also include information freshness through age of information (AoI)~\cite{kaul2012} and its semantic variants, such as Version AoI and age of misclassified information (AoMI)~\cite{EHAoI,SemAware6G,VisualEventLEO}. Finally, \emph{goal} metrics measure the value delivered by the communication system, including task success probability~\cite{IRST}, semantic rate and spectral efficiency~\cite{9763856}, and semantic energy efficiency~\cite{10460718}. The latter is particularly relevant when on-board energy is limited. NTN studies therefore often combine a fidelity or task metric with a cost metric to capture both performance and the resources required to achieve it. However, the wide variety of metrics also makes numerical results across studies difficult to compare directly.

\subsection{Five NTN Constraints and the Corresponding Mechanisms}
\label{subsec:constraints}

NTN links challenge conventional communication in five structurally distinct ways. We formulate each as a constraint and map it to the semantic mechanisms that address it, together with the cost each mechanism introduces. This constraint-mechanism-cost view is summarized in Table~\ref{tab:Mapping} and provides the basis for the design choices discussed below.

\begin{table*}[t]
\centering
\caption{\textcolor{black}{Mapping the Five Structural NTN Constraints to the Semantic Mechanisms That Answer Them. The final column names the cost each mechanism incurs, which is what makes the mapping a design trade-off rather than a list of benefits.}}
\label{tab:Mapping}
\footnotesize
\renewcommand{\arraystretch}{1.1}
\setlength{\tabcolsep}{4pt}
\begin{tabular}{|>{\centering\arraybackslash}p{0.55cm}|p{4cm}|p{6cm}|p{1.8cm}|p{4.0cm}|}
\hline
\textbf{ID} & \textbf{NTN constraint} & \textbf{Semantic mechanism that answers it} & \textbf{Representative works} & \textbf{Cost the mechanism incurs} \\ \hline
C1 & Adverse and time-varying propagation: FSPL $>200$~dB, rain fade, Doppler, $10$-$20$~dB SNR swing per pass & Smooth distortion-SNR behaviour of D-JSCC removes the cliff effect; channel-adaptive rate and decoder selection track the state; generative decoding absorbs residual symbol errors; importance-aware protection preserves task-critical regions & \cite{Bourtsoulatze2019DeepJSCC,CCB,IRST,FMSat,DopplerAdaptive,11161311} & Reconstruction quality becomes a stochastic function of the channel; no fidelity guarantee for archival or scientific data \\ \hline
C2 & Delay and contact scarcity: $5$~ms to $22$~min one-way, $5$-$15$~min LEO pass, stale feedback & Transmit the actionable result rather than its inputs; delta-encode against a shared KB; schedule by semantic freshness and version AoI; semantic-aware DTN bundle prioritization & \cite{Soret2024SemEdge,VisualEventLEO,uysal2024goal,EHAoI,10901032} & The relevance decision is made on-board, before the ground segment sees the data, and cannot be revisited \\ \hline
C3 & SWaP-C limits: tens to hundreds of watts, flight processors $10$-$15$ years behind ground GPUs, UAV compute-versus-endurance & Asymmetric architecture: a lightweight on-board encoder (sparse linear, quantized, distilled, or codebook) with a heavy ground decoder; energy-aware bit/semantic mode switching; offload of extraction to HAPS or peer satellites & \cite{CompressedLearning,Soret2024SemEdge,11207608,11006980,nguyen2026anchor} & Transmitter and receiver can no longer be trained or updated as a single unit \\ \hline
C4 & Multi-hop heterogeneity: per-hop SNR, Doppler, and latency differ by orders of magnitude; distortion accumulates & Per-hop semantic denoising trained with the codec; latent- and codebook-domain forwarding that skips full re-encoding; joint semantic coding and routing optimizing end-to-end fidelity & \cite{Nguyen2024NovelDenoising,LinSemForward,SemForwardCodebook,GraphJSCR,10345598} & Intermediate nodes must hold matched model state, turning routing into model placement \\ \hline
C5 & No trusted single operator: multi-vendor paths, shared KB as shared state, jamming asymmetry against low-redundancy streams & Physical-layer secure SemCom and friendly jamming; differentially private feature transformation; adversarial training across the channel; blockchain-anchored federated training & \cite{MissionSecureSat,nkurunziza2026attention,PairedARN,elmahallawy2025decentralized,Hu2025SAGIN} & Every defence costs either latency, which NTN cannot spare, or accuracy, which the task cannot spare \\ \hline
\end{tabular}
\end{table*}

\textit{C1: Adverse and time-varying propagation:}
Beyond high propagation loss, NTN links also experience channel conditions that can vary substantially over time and across operating environments. Rain, gaseous absorption, scintillation, and deep fading can further reduce the available SNR, with attenuation ranging from several to tens of dB under adverse conditions~\cite{series2015propagation,series2019attenuation,IGWE201952,basarudin2024rain,al2025channel}. Mobility adds another source of variation. A LEO satellite traveling at roughly $7.5$~km/s can produce Doppler shifts of tens to hundreds of kHz. Even after GNSS-aided pre-compensation, the residual Doppler can cause significant inter-carrier interference, particularly when high-order modulation is used. Following the small-offset approximation in~\cite{sathananthan2001}, the Doppler shift and the resulting interference in an orthogonal frequency-division multiplexed waveform can be expressed as:
\begin{equation}
\label{eq:doppler_ici}
f_d = \frac{v_{\text{rel}}}{c}\, f_c\, \cos\vartheta,\qquad \mathrm{ICI}\approx \frac{\pi^2}{3}\left(\frac{f_d}{\Delta f}\right)^2 ,
\end{equation}
where $v_{\text{rel}}$ is the relative velocity, $\vartheta$ the angle between the line of sight and the velocity vector, and $\Delta f$ the subcarrier spacing. Three SemCom mechanisms can help with this challenge. D-JSCC provides a smooth degradation in reconstruction quality as SNR decreases ~\cite{11476190,Bourtsoulatze2019DeepJSCC}; channel-adaptive codecs adjust their transmission strategy according to the measured or predicted channel state~\cite{CCB,IRST}. The generative decoders can recover a plausible source even when the transmitted symbols are corrupted~\cite{FMSat,10960413}. These mechanisms therefore make reconstruction quality depend continuously on channel conditions rather than on a fixed error threshold. This behavior is useful for applications such as monitoring, where approximate reconstruction may still be informative.

\textit{C2: Delay and contact scarcity:}
Round-trip delay ranges from $5$-$12$~ms for LEO to $100$-$250$~ms for MEO and $500$-$700$~ms for GEO~\cite{NTNTutorial}, while a LEO satellite may remain visible for only five to fifteen minutes. Long delays and short contact windows make feedback-based mechanisms less effective. Feedback for hybrid automatic repeat request or adaptive coding may arrive too late, closed-loop control becomes more difficult, and the amount of data that can be transferred is limited by the contact window. SemCom cannot reduce propagation delay, but it can make better use of the available transmission time by prioritizing information that is more useful for the task. Goal-oriented approaches, for example, transmit an actionable result instead of the raw telemetry needed to derive that result at the receiver~\cite{Soret2024SemEdge,VisualEventLEO,IoSSemCom}. Semantic-aware delay/disruption-tolerant networking (DTN) can similarly prioritize bundles based on their relevance~\cite{uysal2024goal}. Other approaches use delta encoding with shared knowledge~\cite{ThomasReason1,Hyowoon,LiuFangyu} or freshness-aware scheduling~\cite{EHAoI,10901032} to spend the limited contact time on information with greater task value. However, these approaches require the transmitter to know which information is important to the receiver.

\textit{C3: SWaP-C limits on the platform:}
Small LEO satellites typically operate with only tens to hundreds of watts of available power, while their radiation-hardened processors can lag commodity ground-based GPUs by roughly ten to fifteen years~\cite{lovelly2017comparative,ortiz2023onboard}. UAVs face a similar constraint because computation and flight share the same energy budget: energy spent on semantic processing reduces the energy available for staying airborne. The computation required by SemCom must therefore be considered as part of the overall energy budget. Semantic processing is beneficial only when its computation cost is smaller than the transmission energy it saves. Otherwise, SemCom may consume more energy than the conventional communication method it is intended to replace~\cite{Soret2024SemEdge}. Transmit power $P$, bandwidth $B$, this trade-off is given as:
\begin{align}
\label{eq:onboard_energy}
E_{\mathrm{total}}(\rho) &= \underbrace{E_{\mathrm{enc}}(\rho)}_{\text{on-board}} + \underbrace{\frac{k(\rho)}{B} P}_{\text{transmission}},\\
\rho^{\star} &= \arg\min_{\rho}\, E_{\mathrm{total}}(\rho)\;\;\text{s.t.}\;\; Q(\rho)\geq Q_{\min},
\end{align}
where $\rho$ controls the compression level. A smaller compression ratio $\rho$ reduces the number of transmitted symbols $k(\rho)$, but it also increases the encoder energy $E_{\mathrm{enc}}(\rho)$. The selected $\rho$ must therefore balance on-board computation and transmission energy while maintaining the required task quality $Q_{\min}$. One practical way to achieve this balance is architectural asymmetry: a lightweight on-board encoder, such as a sparse linear map~\cite{CompressedLearning}, a quantized or distilled network, or a codebook lookup, is paired with a computationally expensive decoder on the ground~\cite{rombach2022high}. Energy-aware switching between bit and semantic modes~\cite{11207608} and computation offloading to HAPS or neighboring satellites~\cite{11006980} provide additional ways to reduce the on-board burden. 

\textit{C4: Multi-hop heterogeneity:}
An end-to-end NTN path may involve several hops, from a user to a UAV or LEO satellite, across one or more ISLs, through a feeder link to a gateway, and finally over terrestrial backhaul. These hops can experience very different SNRs, Doppler shifts, propagation delays, and atmospheric conditions. A semantic transmission strategy that works well on one hop may therefore perform poorly on another. In regenerative relaying, intermediate nodes decode and re-encode the received information, which can accumulate semantic distortion over multiple hops. In contrast, transparent relaying forwards the signal without decoding, allowing channel noise to accumulate along the path~\cite{Nguyen2024NovelDenoising,10345598}. Several approaches address these problems. Per-hop semantic denoising reduces the distortion introduced at each hop by jointly training the denoiser with the semantic codec~\cite{Nguyen2024NovelDenoising}. Semantic-forward relaying instead forwards information directly in the latent or codebook domain, avoiding repeated conversion between semantic and bit representations~\cite{LinSemForward,SemForwardCodebook}. Joint semantic coding and routing takes a path-level approach by optimizing end-to-end semantic fidelity rather than treating each hop independently~\cite{GraphJSCR}.

\textit{C5: Absence of a trusted single operator:}
An NTN path may span multiple organizations, with the satellite, HAPS relay, gateway, and user terminal operated by different entities. This creates a trust challenge for SemCom because a shared KB requires these operators to maintain compatible information or models without necessarily sharing private data or model parameters. Security is also more difficult because NTN links cover large geographic areas, allowing a ground-based jammer to disrupt a large part of the transmission~\cite{Hu2025SAGIN}. Existing solutions include physical-layer secure SemCom~\cite{MissionSecureSat,11011492,10870357}, differentially private feature transformation~\cite{nkurunziza2026attention}, and blockchain-based federated training across multiple operators~\cite{elmahallawy2025decentralized}. These protections, however, come at a cost. Security and privacy mechanisms can increase communication and computation overhead, while stronger protection can also remove information needed for accurate semantic recovery.

\begin{table*}[t]
\centering
\caption{\textcolor{black}{Severity of each constraint by NTN platform: \textbf{H} = high (dominant design driver), \textbf{M} = moderate, \textbf{L} = low.}}
\label{tab:platformvectors}
\footnotesize
\renewcommand{\arraystretch}{1.1}
\setlength{\tabcolsep}{4pt}
\begin{tabular}{|p{1.0cm}|c|c|c|c|c|p{7cm}|p{6cm}|}
\hline
\textbf{Platform} & \textbf{C1} & \textbf{C2} & \textbf{C3} & \textbf{C4}  & \textbf{C5} & \textbf{Mechanisms the profile justifies} & \textbf{Mechanisms of low value here} \\ \hline

GEO & H & H & L & L  & M & Aggressive compression, generative reconstruction from sparse cues, rain-fade mode switching & Channel prediction (geometry is static); contact scheduling \\ \hline

MEO & M & M & M & M  & M & Balanced designs; multi-orbit relay and gateway diversity & Extreme compression (window is not the bottleneck) \\ \hline

LEO & H & M & H & H  & H & Predictive rate control from ephemeris, importance-aware protection, semantic-forward relaying & Static codebooks; any scheme needing stable long-lived per-node state \\ \hline

HAPS & L & L & M & M  & M & On-board heavy decoding, KB hosting and aggregation, semantic caching, cross-tier translation & Cliff-avoidance coding (link is stable); delta encoding against contact windows \\ \hline

UAV & M & L & H & M  & M & Joint trajectory and compression optimization, ultra-light encoders, codebook lookup, semantic forwarding in swarms & Doppler-adaptive waveforms; ephemeris-based prediction \\ \hline

Deep space & H & H & H & H & L & Extreme goal-oriented abstraction, DTN bundle prioritization, quantum-assisted representations & Closed-loop adaptation of any kind; feedback-driven HARQ \\ \hline
\end{tabular}
\end{table*}

\subsection{Platform Constraint Vectors}
\label{subsec:platformvectors}

The five constraints apply to all NTN platforms, but their severity differs across platforms. We describe this combination as the platform's constraint vector. This vector helps determine which semantic mechanisms are practical for a given platform and whether their benefits justify the associated cost. Table~\ref{tab:platformvectors} summarizes these constraint vectors.
\begin{itemize}
    \item A GEO satellite link experiences high path loss and long propagation delay~\cite{9861699}. However, its fixed geometry and relatively large power budget make the channel slow and predictable~\cite{MahboobSatelliteSurvey}. Under these conditions, aggressive compression and generative reconstruction are generally more useful than channel prediction.

    \item LEO operates under almost the opposite conditions: path loss and delay are lower, but the channel changes rapidly, and the contact window is short~\cite{luo2026learning,khalid2024characterization}. Predictive rate control and importance-aware protection are therefore more useful~\cite{IRST}, allowing the system to adapt to upcoming channel changes and prioritize important information before the link is lost.

    \item UAV links have relatively mild propagation conditions and very low delay, but UAVs operate under tight energy constraints~\cite{9420293}. Their trajectories can also be controlled. As a result, many UAV-based SemCom studies jointly optimize trajectory and compression to reduce energy consumption~\cite{10419536,XIE2025103762}, while relatively few focus on channel prediction~\cite{RobustUAVPredictive}.
    
    \item HAPS provides favorable communication and computing conditions~\cite{FLSTRA}, including quasi-stationary links, low propagation delay, and sufficient payload capacity to support dedicated accelerators~\cite{9380673}. Despite these advantages, only a few works have studied SemCom specifically for HAPS.
\end{itemize}

\subsection{The Proposed Taxonomy}
\label{subsec:taxonomy}

We organize the NTN-SemCom literature into five design planes based on where semantic mechanisms operate in the system, together with a trust plane that spans all five. These planes follow the main decisions involved in designing a deployable NTN-SemCom system: what information to represent and where to process it; how to transmit and protect the representation over the link; how to deliver it across a heterogeneous multi-hop network; how to allocate the required communication and computing resources; and how to maintain the shared models, codebooks, and knowledge bases over time.

\begin{itemize}
    \item \emph{Plane 1: Semantic Representation and On-board Encoding} (Section~\ref{sec:Representation}). This plane determines what semantic representation is transmitted and where the encoding process is performed. We first classify works by the form of representation and then by encoder placement under SWaP-C constraints.

    \item \emph{Plane 2: Channel-Adaptive Semantic Transmission} (Section~\ref{sec:ChannelAdaptive}). This plane concerns how the semantic transmission system adapts to changing NTN channel conditions. We classify works according to what part of the system is adapted and what information drives that adaptation.

    \item \emph{Plane 3: Semantic Networking and Multi-Hop Transport} (Section~\ref{sec:Networking}). This plane addresses how semantic content is carried across heterogeneous multi-hop paths. Works are grouped by the network function they perform, including distortion control, forwarding, routing, mobility management, and delay-tolerant scheduling.

    \item \emph{Plane 4: Semantic Resource Management and Cross-Layer Optimization} (Section~\ref{sec:Resource}). This plane considers the communication, computation, and energy resources required by a semantic system and how they should be allocated. We organize the literature according to the source of the resource-management challenge.

    \item \emph{Plane 5: Distributed Learning and Knowledge-Base Management} (Section~\ref{sec:Learning}). This plane addresses how the shared semantic assets required by the system, such as models, codebooks, and knowledge bases, are constructed, distributed, updated, and kept consistent over time. Works are classified according to the type of asset or learning process being distributed.

    \item \emph{Trust Plane: Security, Privacy, and Safety} (Section~\ref{sec:Security}). This plane cuts across the other five because vulnerabilities can arise in one part of the system and affect another. For example, an attacker may target the transmitted representation, compromise an intermediate relay, or poison a distributed model update.

\end{itemize}

\section{Semantic Representation and On-Board Encoding}
\label{sec:Representation}

A fundamental design choice in an NTN SemCom system is what information to transmit over the channel. The choice of representation affects both how the information is transmitted and how it is recovered at the receiver. A continuous latent, for example, must be converted into a form that can be carried by standardized digital waveforms. A generative conditioning signal sends only part of the source information and relies on the receiver to generate the remaining content. A symbolic triplet goes further by relying on shared background knowledge to recover its meaning. Based on what is transmitted, we group existing approaches into five classes: continuous latents, discrete codebook indices, generative conditioning signals, symbolic graphs, and event or significance indicators. Across these classes, the transmitted payload generally decreases, while the receiver's reliance on prior knowledge or shared state increases.

\subsection{Continuous Latent Representations}
\label{subsec:continuous}

The most direct approach maps the source to continuous channel symbols using a learned encoder, following the analog D-JSCC framework in~\cite{Bourtsoulatze2019DeepJSCC}. This approach is particularly relevant to satellite communication because link quality can vary substantially during a pass. Unlike conventional digital transmission, D-JSCC degrades gradually as the channel worsens, avoiding a sharp drop in reconstruction quality. Continuous latent representations can also compress high-dimensional data, such as images, without first converting them into a discrete symbol alphabet.

Designs in this class differ mainly in what the transmitted latent is expected to support at the receiver. Reconstruction-oriented systems preserve enough information to reconstruct the source, without assuming a specific downstream task. Task-oriented systems instead transmit only the information needed for a given task. For example, the wildfire-detection system in~\cite{WildfireUAV} transmits semantic features that are directly classified at the receiver, without reconstructing the original image. The study also shows that detection accuracy does not necessarily improve with UAV altitude: a higher altitude provides a wider field of view but makes the fire signature smaller. In contrast,~\cite{9796572} preserves the block structure of the image and uses DRL to decide which blocks to transmit. This turns the trade-off between task accuracy and transmission latency into an explicit scheduling decision rather than leaving it to the semantic codec alone.

The second question is whether the latent representation is designed for a single link or a multi-satellite sensing task. Most existing approaches encode the observation of each satellite independently. The cognitive semantic augmentation framework for LEO Earth observation~\cite{CSA_LEO} instead combines multispectral observations from multiple satellites in the semantic domain through ISLs before transmitting the result to the ground. This approach is particularly useful for large-scale events, such as wildfires or flood fronts, that extend beyond the footprint of a single satellite. In these cases, combining semantic information from multiple observations can provide information that no individual satellite captures alone. Other studies extend continuous-latent transmission by adapting to distribution shifts online~\cite{10158528} or adjusting the coding strategy to shadowed-Rician fading~\cite{10681686}.

Overall, continuous-latent transmission is well established, but its integration into practical satellite systems remains difficult. Analog latent transmission requires software-defined radio capabilities and high-resolution signal conversion, which may be supported by regenerative payloads but are less suitable for legacy bent-pipe satellites. It is also difficult to integrate analog semantic signals with standardized digital waveforms carrying conventional bit-level traffic. These limitations motivate the use of discrete representations, as discussed next.

\subsection{Discrete and Codebook Representations}
\label{subsec:discrete}

A second class uses a discrete codebook to make semantic transmission compatible with existing satellite waveforms. The encoder quantizes the latent representation into codebook indices, which can then be mapped to digital symbols and transmitted using standardized waveforms such as DVB-S2X or NR-NTN. Digital D$^2$-JSCC~\cite{D2JSCC} provides the basic framework, which later NTN studies adapt to different transmission environments. One direction focuses on adapting the discrete representation to the physical channel. Free-space optical links provide high bandwidth and favorable SWaP characteristics, but atmospheric turbulence and pointing jitter produce channel impairments that differ from the Gaussian noise assumed by many semantic codecs. To address this issue,~\cite{10969992} combines a vector-quantized autoencoder with spatial normalization. Under M\'alaga-distributed turbulence, the proposed system achieves more than $3$~dB of power gain and continues to operate at zenith angles above $60^\circ$, where the conventional system fails. The main benefit is therefore not only improved transmission performance, but also a wider operating range under challenging optical channel conditions.

A second line of work uses the discrete representation to divide semantic information across channels with different capacities. The dual-level architecture in~\cite{11331466} transmits a coarse codebook index over a low-rate telemetry channel and a finer residual over a higher-capacity data channel. This design allows the receiver to recover a basic representation even when only the low-rate channel is available, while the residual progressively improves its quality. As a result, performance at very low rates depends largely on how well the coarse codebook represents the expected scene distribution, rather than on the capacity of the channel alone. A common assumption underlies this entire class but is rarely discussed. Every codebook-based scheme requires the transmitter and receiver to use the same codebook, yet the reviewed works generally assume that this codebook is fixed and distributed in advance. However, the data distribution can change with season, geographic location, or sensor degradation. Keeping the codebook useful and consistent at both ends therefore becomes a maintenance problem, involving updates, synchronization, and drift management, which is discussed in Section~\ref{sec:Learning}.

\subsection{Generative Conditioning Signals}
\label{subsec:generative}

The third class does not transmit the source itself or a latent representation that is directly decoded into the source. Instead, it sends a compact conditioning signal that guides a generative model at the receiver. This approach is particularly useful under the severe channel conditions described by constraint C1. During a deep fade, a conventional decoder may reconstruct a heavily corrupted image, whereas a generative decoder can still produce a plausible and potentially useful output, although some details may differ from the original~\cite{11387839,10628028}. The main trade-off is therefore between fidelity and availability: the generated output may not reproduce the source exactly, but it can remain useful when conventional reconstruction fails. FMSAT~\cite{FMSat} is a representative NTN example. The satellite converts the image into a semantic layout and transmits it to a gateway, where a conditional diffusion model reconstructs the image. The design also considers two characteristics of satellite communication. First, because long propagation delays make retransmission costly, a coarse detector identifies corrupted information on board. Second, because satellites can revisit the same ground track, a previously received clean image is used as prior information to repair regions affected by attenuation. Other designs use different forms of conditioning. Some convert the image into a short textual prompt~\cite{10628028}, whereas others extract semantic features according to the context requested by the receiver rather than using a fixed representation learned during training~\cite{DSCUAV}. Generative conditioning can also extend across multiple modalities. The audiovisual satellite system in~\cite{LiuFangyu}, for example, uses an LLM agent to decide whether audio or video should be prioritized. It also updates keyframes to keep the decoder's reference information aligned with the current source. This illustrates an important requirement of generative SemCom: the transmitter and receiver often rely on shared state, which must remain synchronized as the source changes.

Two limitations appear repeatedly across this class but are rarely evaluated directly. First, generative decoding can be computationally expensive. For diffusion-based decoders, the cost increases with the number of denoising steps, making the receiver's computing capacity part of the communication design itself. This coupling is explicitly considered in~\cite{huang2026semantic}, which jointly optimizes transmission and denoising depth. Second, generative reconstruction does not guarantee that every generated detail is faithful to the original source. The perception-error analysis in~\cite{10960413} examines this issue, but the problem goes beyond conventional rate-distortion performance because a perceptually plausible output may still contain details that were not present in the source. An inaccurate detail may be acceptable if the output remains useful for its intended task, but it becomes problematic when the model generates information that was not present in the source, and that information affects a downstream decision. We discuss this distinction further in Section~\ref{subsec:safety}.

\subsection{Symbolic and Graph-Structured Representations}
\label{subsec:symbolic}

A fourth class transmits structured symbolic information rather than an image or latent representation. The transmitter represents a scene as (head, relation, tail) triplets and sends these triplets to the receiver, which uses a shared knowledge base to recover the full meaning~\cite{ThomasReason1,Hyowoon,ChristinaLessdata}. This approach offers two main benefits. First, it can greatly reduce the amount of transmitted data, since a scene containing only a few objects can be represented by a few hundred bytes. Second, the graph structure provides some robustness to transmission errors. If one entity is corrupted, the receiver may still recover it from its relations with the remaining entities. A representative NTN example is the UAV cognitive semantic communication system in~\cite{KGSemUAV}, which targets object detection under limited computation, energy, and bandwidth. Rather than reconstructing a compressed image, the receiver combines knowledge-graph information with visual proposals extracted from the image and refines the result using a relational graph attention network. By reasoning over the relations between objects, the network can correct errors caused by both compression and channel impairments. On a real aerial-image dataset, the system reports only a $0.6\%$ accuracy loss under Rayleigh fading, compared with $7.4\%$ for a JPEG-plus-LDPC baseline. This result shows that the main gain comes from reasoning over the structured representation, rather than from the codec alone.

Other works relax two assumptions made by earlier approaches: that all users require the same content and that the extracted graph is free from errors. The personalized-saliency framework in~\cite{9959884} addresses the first assumption by selecting different triplets for different users based on objective attention and user-specific heatmaps. It also analyzes the probability of triplet loss directly under Fisher-Snedecor $F$ fading, rather than treating the channel as a black box. In contrast,~\cite{10872947} focuses on robustness by applying adversarial training to the graph construction process, making the extracted triplets more resistant to semantic noise. A third direction considers applications that naturally use structured information. In~\cite{10750853}, a UAV transmits scene graphs to update a digital twin, which can directly use entities and their relations instead of receiving the original pixels. A common requirement across these approaches is that the transmitter and receiver share the same knowledge base. However, the reviewed works do not examine what happens when the two knowledge bases become inconsistent. This is particularly important in NTNs, where communication windows can last only a few minutes, and data may pass through multiple administrative domains. Knowledge-base synchronization therefore cannot be assumed. If the transmitter and receiver use different knowledge, the received semantic representation may be interpreted incorrectly or may no longer support reliable reconstruction.

\subsection{Event and Significance Representations}
\label{subsec:event}

Event and significance representations take a different approach from the previous four classes. Rather than finding a more compact representation of an observation, they first decide whether the observation is important enough to transmit. The main question therefore shifts from \emph{how to represent the observation} to \emph{whether it should be transmitted at all}. A clear EO example is the change-detection approach in~\cite{10659182}, which exploits the fact that much of a multispectral image remains unchanged between successive passes. The method removes cloud-obscured regions, estimates the change likelihood of each pixel, and transmits only pixels with high change scores. The receiver reconstructs the observation using a stored reference image, with reported energy savings of up to $58\%$. Removing cloud-covered regions is particularly useful because transmitting them consumes contact time and energy while providing little useful information for the downstream task. While~\cite{10659182} evaluates the importance of individual pixels,~\cite{VisualEventLEO} identifies and prioritizes events. It uses AoMI as the main metric instead of reconstruction quality, shifting the objective from accurately reconstructing the source to delivering important event information in a timely manner. A similar idea appears in task-specific detectors such as~\cite{10256109}, which transmit only regions containing detected objects and use detector confidence to determine their importance. Significance-based transmission also extends beyond images to status updates. For example,~\cite{EHAoI} controls when a device sends an update to minimize version AoI, whereas~\cite{SemAware6G} considers freshness, relevance, and utility across multiple network tiers. Across these approaches, the main design decision is \emph{what} and \emph{when} to transmit, rather than how to encode the entire source.

\subsection{Encoder Placement under SWaP-C Constraints}
\label{subsec:placement}

The classes above describe \emph{what} is transmitted; this subsection considers \emph{where} semantic encoding is performed. Unlike terrestrial systems, NTN platforms operate with limited compute, memory, power, and hardware resources, as discussed under constraint C3. Encoder design must therefore balance representation quality against on-board feasibility. One approach is \emph{on-board lightweight encoding}, which reduces encoder complexity to fit within flight-hardware constraints. In~\cite{CompressedLearning}, the deep encoder is replaced by a learned sparse compression matrix, reducing on-board processing to sparse linear operations that can run on a radiation-hardened FPGA. Although a deep encoder could provide better compression, its higher computational cost can make on-board deployment impractical. Encoder performance should therefore be evaluated together with its hardware cost: gains in representation quality have limited practical value if the encoder cannot operate within the platform's SWaP-C constraints.

\emph{Asymmetric encoder-decoder design} keeps a learned encoder but assigns different computational loads to the transmitter and receiver. In~\cite{nguyen2026anchor}, the authors consider heterogeneous user devices and hint at the potential to shift more computation to the receiver. Generative SemCom follows a similar asymmetric design: the transmitter sends a compact representation, while a more computationally intensive generative model reconstructs the source at the receiver~\cite{FMSat,rombach2022high}.

\emph{Offloading and splitting} take this idea further by moving part or all of the semantic processing away from the constrained platform. Soret \emph{et al.}~\cite{Soret2024SemEdge} propose a distributed LEO edge architecture that splits processing between on-board feature extraction and edge-side computation. The split is selected by balancing transmission and computation energy. As shown in Eq.~\eqref{eq:onboard_energy}, the best split therefore depends on both the processor architecture and the current channel conditions. When the constrained node cannot perform semantic extraction locally, the computation can instead be offloaded to a HAPS or another satellite~\cite{11006980,11310140}, or shared across multiple nodes through split learning~\cite{10445211}.

\begin{table*}[!t]
\centering
\caption{\textcolor{black}{Summary of Semantic Representation and On-Board Encoding Approaches for NTN.}}
\label{tab:representation}
\scriptsize
\renewcommand{\arraystretch}{1.1}
\setlength{\tabcolsep}{2.6pt}
\begin{tabular}{|>{\centering\arraybackslash}p{1.45cm}|>{\centering\arraybackslash}p{0.75cm}|c|c|>{\centering\arraybackslash}p{1.5cm}|>{\centering\arraybackslash}p{2.6cm}|>{\centering\arraybackslash}p{1.9cm}|p{6.9cm}|}
\hline
\textbf{Class} & \textbf{Ref.} & \textbf{Plat.} & \textbf{Par.} & \textbf{Modality} & \textbf{Representation transmitted} & \textbf{Reported metric} & \multicolumn{1}{c|}{\textbf{Mechanism and principal limitation}} \\ \hline

\multirow{5}{1.45cm}{\centering\textbf{Continuous latents}}
& \cite{WildfireUAV} & U & DJ & Aerial image & Continuous JSCC features & Detection accuracy, PSNR & Classifies on received features, skipping reconstruction. Single-task encoder; accuracy varies non-monotonically with altitude. \\ \cline{2-8}
& \cite{CSA_LEO} & S & DJ & Multispectral & Discrete task-oriented JSCC codes & Detection accuracy & Routing satellite fuses multi-satellite observations over ISLs before downlink. Requires matched encoders constellation-wide. \\ \cline{2-8}
& \cite{9796572} & U & DJ & Image blocks & Selected image-block latents & Accuracy, latency & DRL selects transmitted blocks under a joint accuracy-latency reward. Block granularity fixed a priori. \\ \cline{2-8}
& \cite{10681686} & S & DJ & Image & Rate-controlled latents & PSNR & Adaptive coding under shadowed-Rician fading. Rate policy reactive to measured SNR only. \\ \cline{2-8}
& \cite{10158528} & \xmark & DJ & Image & Online-adapted nonlinear transform codes & Rate-distortion & Online nonlinear transform coding tracks source distribution shift. Not evaluated under orbital dynamics. \\ \hline

\multirow{3}{1.45cm}{\centering\textbf{Discrete / codebook}}
& \cite{10969992} & S & DJ & Remote-sensing image & VQ-VAE indices over FSO & Power gain, zenith angle & Spatial-normalized VQ-VAE over M\'alaga turbulence; $>3$~dB gain, usable beyond $60^\circ$ zenith. Static codebook. \\ \cline{2-8}
& \cite{11331466} & U & DJ & Image & Coarse codeword index + fine residual & Rate, reconstruction & Coarse index on telemetry, fine residual on data channel. Codebook must match the scene distribution. \\ \cline{2-8}
& \cite{D2JSCC} & \xmark & DJ & Image & Quantized digital symbols & Rate-distortion & Digital D-JSCC compatible with standardized modulation. Quantization and channel noise must be co-trained. \\ \hline

\multirow{5}{1.45cm}{\centering\textbf{Generative conditioning}}
& \cite{FMSat} & S & GAI & Remote-sensing image & Foundation-model segmentation layout & Perceptual loss, SSIM & Satellite segmentation, gateway diffusion, split error detection, prior pass as repair reference. Gateway compute unbounded. \\ \cline{2-8}
& \cite{10628028} & \xmark & GAI & Image & Text prompt from a generative KB & Bandwidth, quality & Image reduced to a prompt against a generative KB. No fidelity guarantee on synthesized detail. \\ \cline{2-8}
& \cite{DSCUAV} & U & GAI & Image & ViT features + context prompt & Semantic SSIM, AoI & Prompt-aware encoding makes features context-dependent; 22\% SSS gain, 14\% AoI reduction. Context set predefined. \\ \cline{2-8}
& \cite{LiuFangyu} & S & GAI & Audio + video & Modality-prioritized embeddings + keyframes & Bandwidth, keyframe distance & LLM agent prioritizes audio or video; keyframe update resynchronizes the decoder. Agent inference cost unaccounted. \\ \cline{2-8}
& \cite{10960413} & \xmark & GAI & Image & Foundation-model latents & Perception error & Relates generative reconstruction error to semantic-aware power allocation. Does not address fabricated detail. \\ \hline

\multirow{4}{1.45cm}{\centering\textbf{Symbolic \& graph}}
& \cite{KGSemUAV} & U & KG & Aerial image & Compressed features + KG triplets & Detection accuracy,complexity & Multi-scale codec plus R-GAT reasoning over KG priors; 0.6\% versus 7.4\% accuracy drop under Rayleigh fading. Assumes an aligned KB. \\ \cline{2-8}
& \cite{9959884} & U & KG & Image & Per-user selected RelTR triplets & Task accuracy & Objective saliency fused with per-user heatmaps; triplet loss derived under Fisher-Snedecor $F$ fading. Needs per-user profiles. \\ \cline{2-8}
& \cite{10872947} & U & KG & Image & Caption-derived triplets & Robustness to semantic noise & ViT patches to GPT-2 caption to triplets, hardened adversarially. Caption stage adds latency and a hallucination surface. \\ \cline{2-8}
& \cite{10750853} & U & KG & Image & Scene graphs & Latency (min-max) & Scene graphs sent for digital-twin update from a UAV that is also an edge server. Graph schema task-specific. \\ \hline

\multirow{4}{1.45cm}{\centering\textbf{Event \& significance}}
& \cite{10659182} & S & SIG & Multispectral & Changed pixels after cloud filtering & Energy ($58\%$ saving) & Cloud filtering then per-pixel change scoring against a stored reference. Needs an accurate co-registered reference. \\ \cline{2-8}
& \cite{VisualEventLEO} & S & SIG & Image & Event-classification features & AoMI, accuracy & Sends only event-classification features; AoI analysis gives the fraction of users meeting timeliness. Event vocabulary fixed. \\ \cline{2-8}
& \cite{10256109} & U & SIG & Image & Detected object regions & Transmission cost, quality & YOLO selects regions and detector confidence sets importance. Confidence is a poor proxy for task value. \\ \cline{2-8}
& \cite{EHAoI} & S & SIG & Status updates & Update decision & Version AoI, energy & Energy-harvesting device times updates to minimize version AoI. Contribution is the transmission decision, not the encoder. \\ \hline

\multirow{4}{1.45cm}{\centering\textbf{Encoder placement}}
& \cite{CompressedLearning} & S & DJ & Remote-sensing image & Sparse linear projections & Reconstruction quality & Sparse matrix replaces the deep encoder, so extraction is one matrix-vector product on rad-hard hardware. Compresses worse by design. \\ \cline{2-8}
& \cite{Soret2024SemEdge} & S & DJ & EO imagery & Task-dependent features & Energy split & Distributed LEO edge; states the extraction-versus-transmission energy trade-off as the design equation. No flight-hardware validation. \\ \cline{2-8}
& \cite{nguyen2026anchor} & \xmark & DJ & Image & Shared latents, per-user decoders & Multi-user quality & Heterogeneity placed in ground-side decoders rather than the on-board encoder. Decoder bank grows with user diversity. \\ \cline{2-8}
& \cite{10445211} & S & DJ & Imagery & Split-layer activations & Accuracy, training cost & Pruning-split federated learning partitions the coder between satellite and ground. Split point fixed, not link-adaptive. \\ \hline

\end{tabular}
\end{table*}

Across Table~\ref{tab:representation}, the five classes form a spectrum from transmitting detailed continuous representations to transmitting only events or significance indicators, where G denotes integrated SAGIN, D denotes deep space, and SIG denotes significance-based transmission. Moving along this spectrum reduces the amount of information sent over the channel, but also increases the receiver's reliance on prior knowledge or shared state. Three patterns emerge. First, compatibility with existing digital communication systems is often as important as reducing the transmitted information. Discrete representations mainly address this issue by bridging learned analog representations and digital communication systems. Second, all classes beyond continuous latents rely on some form of receiver-side state, such as a codebook, generative model, knowledge graph, or reference image. Yet the reviewed works do not examine how stale or inconsistent receiver-side state affects their reported gains. Third, although encoder placement is widely recognized as an important design choice, few studies evaluate it under practical flight-hardware constraints, with~\cite{CompressedLearning} being a notable exception.

\section{Channel-Adaptive Semantic Transmission over NTN Links}
\label{sec:ChannelAdaptive}

Having determined what to transmit, the next question is how the semantic representation should adapt as the channel changes. This is particularly important in NTNs, where channel conditions can vary rapidly during a satellite pass. For example, a LEO link can experience a $10$-$20$~dB change in SNR and a complete Doppler reversal within minutes. However, because the satellite follows a predictable trajectory, much of this channel variation can be anticipated from ephemeris data. Based on what is adapted and the information used for adaptation, we group existing approaches into six categories: rate adaptation, importance-aware protection, Doppler-aware modulation \& waveform design, proactive adaptation using predicted channel conditions, switching between semantic and bit-level transmission, and adaptation for optical \& deep-space.

\subsection{Rate Adaptation to Time-Varying SNR}
\label{subsec:rateadapt}

The most direct way to adapt to a changing link is to adjust the compression ratio: transmit more channel symbols per source symbol when the channel is weak and fewer when it is strong. Existing designs differ mainly in the channel information they use and the constraint they optimize. The simplest approach relies only on the measured SNR. For example, the satellite-integrated Internet design in~\cite{10681686} uses a rate-control policy network to adapt transmission over a shadowed-Rician channel. This provides a basic reactive strategy, but its decision is based on channel information that may already be outdated when the adaptation takes effect. SatSemCom~\cite{11161311} addresses this limitation by placing a channel predictor before the codec, using predicted rather than only measured channel conditions. The underlying idea is that improving channel-state prediction can be more effective than changing the adaptation policy itself when channel variation is rapid. 

Both designs focus on the channel state but do not account for the on-board transmission queue. This can be problematic for EO satellites since an encoder may generate images faster than the link can transmit them, causing the queue to grow during the contact window. The adaptive rate-control framework of~\cite{luo2026learning} addresses this issue by including the transmission queue in the system state and selecting the encoder's channel dimension to maximize the number of images that satisfy a given quality constraint within a finite visibility window. Its objective is therefore defined at the mission level: the number of images delivered above a quality threshold per pass-rather than at the individual image level. This is more appropriate when the contact window is the main resource constraint. The framework also performs SNR prediction at the ground gateway instead of on board, taking advantage of the gateway's greater computational capacity and unlimited energy.

\subsection{Importance-Aware and Unequal Semantic Protection}
\label{subsec:importance}

Rate adaptation treats all semantic information equally. A second category instead allocates bandwidth and protection according to how important each part of the representation is to the downstream task. The IRST framework of Cao \emph{et al.}~\cite{IRST} provides a representative satellite example. It ranks image regions based on their task relevance and channel conditions, and drops less important regions when bandwidth is limited. Compared with a fixed-rate D-JSCC baseline, its advantage becomes larger at low SNR, showing that stronger protection of task-relevant content is particularly useful under poor channel conditions. A related UAV framework~\cite{RobustUAVPredictive} prioritizes information in a different way. Instead of ranking spatial regions, it separates an aerial image into a structural component that captures geometry and a stochastic component that captures texture. When bandwidth is limited, the structural component is transmitted with higher priority, while the receiver reconstructs the texture using a conditional generative prior. The authors report a $5.6$~dB PSNR gain. The two approaches therefore differ in what they prioritize: IRST selects important \emph{regions}, whereas~\cite{RobustUAVPredictive} prioritizes important \emph{properties} of the signal. The latter reduces the amount of information that must be transmitted by generating less critical texture at the receiver while preserving the structural information that cannot be reliably generated.

\subsection{Doppler-Aware Coding, Modulation, and Waveform Design}
\label{subsec:doppler}

Doppler is an important NTN impairment since it changes continuously as the satellite moves. The works that address Doppler directly therefore introduce adaptation mechanisms that are less common in terrestrial networks. At the modulation layer,~\cite{DopplerAdaptive} shows that fixed SNR thresholds are not sufficient for modulation selection when the frequency offset changes throughout a satellite pass. The proposed DRL agent selects the modulation scheme based on predicted Doppler and delay spread. A vector-quantized autoencoder produces discrete codes that can be transmitted through a conventional digital communication pipeline, making the approach suitable for constrained terminals that cannot support analog semantic transmission. At the waveform layer, OTFSSC~\cite{11423904} addresses Doppler by adapting the waveform rather than the modulation scheme. It maps semantic features onto a delay-Doppler grid to reduce inter-carrier and inter-symbol interference, while a diffusion model corrects the remaining distortion. The framework also supports users with different mobility levels without requiring successive interference cancellation. It is evaluated using NTN-TDL-A and vehicular channels, making it one of the few studies in this literature to use a standardized NTN channel model rather than only an additive Gaussian channel. A third approach uses Doppler estimation as part of the semantic task itself. The multi-task framework in~\cite{ahmadi2025semantic} jointly performs Doppler regression, Doppler-bin classification, and semantic motion classification, reporting up to a $35\%$ reduction in normalized mean-squared error. It also provides calibrated uncertainty estimates that indicate the reliability of the Doppler prediction. An adaptation policy could use this uncertainty to select transmission parameters more conservatively when the prediction is less reliable, rather than relying only on a single estimate of the channel state.

\subsection{Predictive and Ephemeris-Driven Proactive Adaptation}
\label{subsec:predictive}

The preceding approaches adapt to channel conditions after they have been observed, which creates a delay between channel measurement and adaptation. NTN provides an opportunity to reduce this delay since satellite motion is predictable. Elevation, range, path loss, and Doppler over a satellite pass can be estimated from ephemeris information before the pass begins. A few studies use channel prediction to support proactive adaptation. The channel predictor in~\cite{11161311} and the SNR-prediction agent in~\cite{luo2026learning} estimate near-term channel conditions instead of relying only on current measurements. The predictive scheduler in~\cite{RobustUAVPredictive} goes a step further by deciding \emph{which content} to transmit in each time slot. It prioritizes structural content when good channel conditions are predicted and defers texture that can later be reconstructed by the generative decoder. Prediction can therefore guide not only the transmission rate but also the order in which semantic content is transmitted. However, these studies mainly predict future conditions from recent channel observations rather than directly using ephemeris information. Exploiting the predictable orbital geometry therefore remains an important direction for NTN-specific adaptation.

\begin{table*}[!t]
\centering
\caption{\textcolor{black}{Summary of Channel-Adaptive Semantic Transmission Approaches for NTN.}}
\label{tab:channel}
\scriptsize
\renewcommand{\arraystretch}{1.1}
\setlength{\tabcolsep}{2.6pt}
\begin{tabular}{|>{\centering\arraybackslash}p{1.55cm}|>{\centering\arraybackslash}p{0.8cm}|c|c|>{\centering\arraybackslash}p{2.3cm}|>{\centering\arraybackslash}p{1.95cm}|>{\centering\arraybackslash}p{2cm}|p{6.5cm}|}
\hline
\textbf{Category} & \textbf{Ref.} & \textbf{Plat.} & \textbf{Par.} & \textbf{What is adapted} & \textbf{Adaptation signal} & \textbf{Reported metric} & \multicolumn{1}{c|}{\textbf{Mechanism and principal limitation}} \\ \hline

\multirow{3}{1.55cm}{\centering\textbf{Rate adaptation}}
& \cite{10681686} & S & DJ & Compression rate & Measured SNR (reactive) & PSNR & Rate-control policy network over shadowed-Rician fading. Always one control interval behind the channel. \\ \cline{2-8}
& \cite{11161311} & S & DJ & Coding strategy, rate & Predicted CSI (proactive) & PSNR, SSIM & Channel predictor placed ahead of the codec to counter stale CSI. Prediction learned from history, not ephemeris. \\ \cline{2-8}
& \cite{luo2026learning} & S & DJ & Channel dimension of Swin-JSCC & Predicted SNR + queue state & Images delivered above quality floor per pass & RL maximizes images meeting quality constraints per visibility window, with queue penalties and gateway-side SNR prediction. Single satellite and gateway. \\ \hline

\multirow{2}{1.55cm}{\centering\textbf{Importance-aware protection}}
& \cite{IRST} & S & DJ & Which regions are sent and how protected & Real-time CSI + task relevance & Accuracy on target regions & Task-driven region selection through a stacked SNR-aware codec, dropping low-importance regions. Region importance is task-fixed. \\ \cline{2-8}
& \cite{RobustUAVPredictive} & U & GAI & Structure vs texture protection and slot order & Predicted A2G link quality & PSNR ($+5.6$~dB), SSIM & Structure-texture VAE prioritizes geometry in good slots and synthesizes texture in bad ones. Prediction uncertainty not propagated. \\ \hline

\multirow{3}{1.55cm}{\centering\textbf{Doppler-aware coding}}
& \cite{DopplerAdaptive} & S & DJ & Modulation order & Predicted Doppler + delay spread & Throughput, accuracy & DRL selects modulation from predicted Doppler; VQ-VAE codes suit digital pipelines. Fixed codebook across Doppler regimes. \\ \cline{2-8}
& \cite{11423904} & G & DJ+GAI & Waveform (delay-Doppler mapping) & Mobility class & Reconstruction, rate region & Features mapped to delay-Doppler grids; diffusion corrects residual distortion; semi-OTFS access without SIC. Receiver complexity unreported. \\ \cline{2-8}
& \cite{ahmadi2025semantic} & S & DJ & Doppler estimate itself & Received waveform & NMSE (-$35\%$), calibrated uncertainty & Joint Doppler regression, binning, and motion classification with calibrated uncertainty. Estimator not closed with any adaptation policy. \\ \hline

\multirow{3}{1.55cm}{\centering\textbf{Hybrid mode switching}}
& \cite{11207608} & G & GAI & Transmission mode (bit / text / text+embedding) & Channel, task, UAV geometry & Semantic communication efficiency & DSAC jointly selects mode, trajectory, pairing, and ISL, with latent diffusion as the highest-compression mode. Mode set hand-designed. \\ \cline{2-8}
& \cite{huang2026semantic} & S & GAI & Mode + receiver denoising depth & Link budget, compute budget & Adaptive-weighted semantic efficiency & Mode, association, ISL migration, and denoising depth optimized jointly. Tunable metric weights risk improvement by reweighting. \\ \cline{2-8}
& \cite{AerialSemRelay} & G & DJ & Per-user mode at the relay & User capability & Sum rate, spectral efficiency & UAV forwards semantic symbols to capable users and transcodes for the rest. Transcoding cost at the UAV not modelled. \\ \hline

\multirow{2}{1.55cm}{\centering\textbf{Optical \& deep space}}
& \cite{10969992} & S & DJ & Codec for turbulent optical channel & Turbulence regime & Power gain, usable zenith angle & VQ-VAE over M\'alaga turbulence; gain expressed as availability rather than rate. Single link, no pointing dynamics. \\ \cline{2-8}
& \cite{uysal2024goal} & D & SIG & What is sampled and when it is forwarded & Contact schedule, information value & Information relevance, delivery & Goal-oriented sampling, grant-free access, and contact-aware prioritization. Framework-level; no codec instantiation. \\ \hline
\end{tabular}
\end{table*}
\subsection{Hybrid Bit-Semantic Mode Switching}
\label{subsec:hybridmode}

A fifth category treats the transmission mode itself as an adaptation variable. The system switches between bit-level and semantic transmission based on channel conditions, task requirements, and available computing resources. This approach is practical because semantic transmission is not always the best choice; its advantages are most relevant when bandwidth is limited or channel conditions are poor. Huang \emph{et al.}~\cite{11207608} demonstrate this approach in a three-tier SAGIN for downlink image delivery. The system switches among bit-level JPEG2000, text-only generation, and text-plus-embedding latent diffusion, while jointly optimizing the transmission mode, UAV trajectories, user pairing, and ISL selection according to reconstruction quality and delay. For direct satellite-to-user access,~\cite{huang2026semantic} extends this idea by jointly selecting the transmission mode and the number of diffusion denoising steps at the receiver. Receiver computation therefore becomes part of the adaptation decision rather than a fixed property of the decoder. Mode switching can also support users with different decoding capabilities. In~\cite{AerialSemRelay}, a UAV forwards semantic symbols directly to semantic-capable users and converts them into bit streams for conventional users. This allows semantic transmission to be used over the satellite-to-UAV bottleneck while maintaining compatibility with conventional terminals.

\subsection{Optical and Deep-Space Regimes}
\label{subsec:optical}

Two regimes require different treatment from the fading- and Doppler-dominated RF links discussed above. In free-space optical (FSO) links, the main impairments are atmospheric turbulence, beam wander, and pointing errors rather than multipath fading. As a result, maintaining link availability can be more important than increasing throughput. The semantic FSO design in~\cite{10969992}, the main NTN example in this regime, follows this direction by extending the range of usable zenith angles rather than maximizing throughput. Deep-space communication represents a more extreme NTN regime, characterized by long propagation delays, infrequent and scheduled contacts, and severe photon constraints~\cite{moision2003deep}. With one-way delays of several minutes, timely receiver feedback is not available for closed-loop adaptation. Uysal~\cite{uysal2024goal} therefore focuses on whether the transmitted information will still be useful when it reaches the receiver. The framework combines goal-oriented sampling, grant-free access, and contact-aware prioritization. In this setting, adaptation is driven by the expected value and timeliness of the information rather than by the instantaneous channel state.

As Table~\ref{tab:channel} shows, these schemes follow the same basic process: they observe or predict the channel state, select a suitable codec configuration, and apply it. The ``adaptation signal'' column separates reactive designs, which respond to the current channel state, from proactive designs, which adapt based on a predicted or known future state. Predictive designs generally report larger gains~\cite{11161311,luo2026learning,RobustUAVPredictive}, particularly for fast-moving platforms. This is because the channel may change significantly between the time it is measured and the time the selected configuration is applied. Two limitations remain. First, existing predictive designs estimate future channel conditions mainly from recent channel observations. However, much of the channel variation in NTNs is determined by orbital geometry and can be predicted well in advance from orbital information. Current approaches therefore learn short-term channel variations that could, in part, be computed directly. Second, most studies consider adaptation only within a short period of a satellite pass. Among the reviewed works, only~\cite{luo2026learning} optimizes adaptation over the entire visibility window. Evaluation is also limited: only~\cite{11423904} uses a standardized NTN channel model. It therefore remains unclear whether the reported gains hold under the channel conditions specified in 3GPP TR 38.811.

\section{Semantic Networking and Multi-Hop Transport}
\label{sec:Networking}

A semantic representation that survives a single link may still degrade over an end-to-end NTN path. Such a path can include an access link, one or more ISLs, a feeder link, and terrestrial backhaul, with large differences in SNR, Doppler, latency, and atmospheric conditions across hops. This creates three problems that do not arise in a single-link system: semantic distortion can accumulate whenever the representation is reconstructed at an intermediate node, relays may need to forward representations they cannot interpret, and the route that maximizes bit throughput may not maximize end-to-end semantic fidelity. We therefore organize the existing work according to the network function modified by the semantic mechanism: distortion control, latent- and codebook-domain forwarding, relay- and gateway-assisted enhancement, semantic-aware routing, mobility management, and finally delay-tolerant scheduling.

\subsection{Distortion Accumulation in Multi-Hop Semantic Chains}
\label{subsec:distortion}

Learning-based semantic codecs are inherently lossy, and this loss can accumulate over multiple hops. When a reconstructed signal is encoded and decoded again, additional information is lost, causing distortion to increase with each hop. Most SemCom systems are designed for a single link and do not address this issue, despite the multi-hop and multi-tier structure of NTNs. The work in~\cite{10345598} directly addresses this problem through recursive training. During training, previously degraded images are fed back into the encoder and decoder, allowing the codec to learn from the degraded inputs that may occur at downstream hops rather than only from clean source data. This approach is simple and computationally inexpensive, providing a useful baseline for NTN-specific solutions. However, recursive training can only limit the accumulated distortion, not eliminate it. Information is still lost each time the signal is encoded and decoded. A more direct solution is therefore to avoid repeated reconstruction by forwarding the semantic representation between intermediate nodes. Alternatively, nodes with sufficient computing resources can enhance the received representation before forwarding it. The choice between these approaches depends on where the processing is performed. Ground gateways can support relatively complex enhancement methods, whereas HAPS operate with a more limited payload budget. Small satellites face the tightest constraints and may therefore favor simple semantic forwarding over additional processing.

\subsection{Latent- and Codebook-Domain Semantic Forwarding}
\label{subsec:forwarding}

Semantic-forward relaying avoids reconstructing the semantic representation at intermediate nodes. Because the relay directly forwards the representation, it avoids additional reconstruction distortion and requires little computation beyond the forwarding operation. Existing designs mainly differ in the type of representation being forwarded. Lin \emph{et al.}~\cite{LinSemForward} introduced a general framework in which the relay extracts and forwards semantic information instead of performing full decode-and-forward. This work establishes semantic relaying as a general cooperative communication mechanism rather than a satellite-specific technique. For LEO satellite-terrestrial systems,~\cite{SemForwardCodebook} further reduces relay overhead by operating directly on discrete codebook indices. The relay can forward or modify these indices without reconstructing the original content, thereby avoiding both regeneration distortion and the associated on-board computation. These approaches assume that the nodes along the path use compatible codecs or share the same codebook. When receivers have different decoding capabilities, however, the relay must translate the representation rather than simply forward it. The dual-mode aerial relay in~\cite{AerialSemRelay} addresses this case by forwarding semantic symbols to users that can decode them and converting them into bit streams for users that cannot. This flexibility comes at the cost of additional computation at the UAV.

\subsection{Gateway- and Relay-Assisted Semantic Enhancement}
\label{subsec:gateway}

A complementary approach is to perform additional semantic processing at an intermediate node with greater computational capacity. Nguyen \emph{et al.}~\cite{Nguyen2024NovelDenoising} consider a constellation serving users with different link qualities. Users near the beam edge experience lower SNR and therefore poorer reconstruction quality, while the gateway receives the same transmission at a much higher SNR. To improve performance for these users, the framework routes their traffic through a gateway, where a lightweight neural network denoises the received latent representation. The gateway is jointly selected as part of this process. Performing denoising directly in the latent domain reduces computational cost because the network processes a compact representation without first reconstructing the image. The authors report higher semantic similarity than direct relaying. The same principle applies to multi-tier NTN architectures. In the satellite-to-HAPS-to-ground downlink of~\cite{10901128}, the intermediate HAPS jointly allocates computation, power, and bandwidth while deciding which semantic information to forward. Semantic enhancement at intermediate nodes therefore becomes part of the broader resource-allocation problem, which we discuss further in Section~\ref{sec:Resource}.

\subsection{Semantic-Aware Routing and Path Selection}
\label{subsec:routing}

Classical satellite routing typically minimizes hop count, delay, or congestion over a time-varying topology~\cite{8425499,8761611,9149175}. These objectives are insufficient for semantic traffic because the quality of the final information depends not only on the selected path but also on the processing performed at each hop and the amount of information transmitted. Optimizing these factors separately can therefore lead to suboptimal end-to-end performance. GraphJSCR~\cite{GraphJSCR} addresses this issue by optimizing them jointly and modeling the constellation as a time-varying directed graph.
\begin{equation}
    \mathcal{G}(t) = \left( \mathcal{V}, \mathcal{E}(t) \right),
\end{equation}
with $\mathcal{V}$ the satellite set and $\mathcal{E}(t)$ the set of active ISLs at time $t$, while forwarding decisions at each satellite $i$ are represented as factorized actions.
\begin{equation}
    \mathbf{a}_i(t) = \left( a_i^{\mathrm{hop}}(t),\ a_i^{c}(t),\ a_i^{\mathrm{relay}}(t) \right),
\end{equation}
where $a_i^{\mathrm{hop}}$ selects the next hop, $a_i^{c}$ allocates the semantic channel budget, and $a_i^{\mathrm{relay}}$ determines whether relay-side re-encoding is used. The authors address the resulting partially observable problem with a graph-attention reinforcement learning policy that uses only local topology and transmission state. This local information makes the approach suitable for constellation-scale operation, where no individual satellite has a complete view of the network. The method also achieves the lowest end-to-end latency among the compared schemes.

The second reason is structural: semantic routing may depend on capabilities that are not available at every satellite. Guo \emph{et al.}~\cite{10740143} consider a mega-constellation in which only some satellites have the AI hardware required for semantic encoding. A feasible semantic path must therefore include at least one satellite with the required encoder and a knowledge base compatible with the destination ground station. Conventional contact-graph routing considers only connectivity and contact windows, so it cannot account for these semantic requirements. To address this limitation, the authors formulate semantic-aware routing as an integer linear program that includes knowledge-base matching constraints and develop a polynomial-time algorithm. Starlink-scale simulations show a higher acceptance ratio and throughput, as well as lower delay, than conventional contact-graph routing. These results show that when semantic processing is available only at selected nodes, routing must consider both the communication path and the semantic capabilities of the nodes along that path. A third approach treats routing as an intent-management problem rather than only a path-selection problem. Gao \emph{et al.}~\cite{10716597} propose a SAGIN architecture in which LLM-driven nodes combine task-driven semantic signaling, semantic routing, and intent-driven cross-domain management. The goal is to shift the network objective from delivering content to completing the intended task. The architecture remains conceptual and is not evaluated against the routing methods discussed above. However, it points to a broader role for semantics: beyond representing the information being transmitted, semantics can also provide a common language for expressing and coordinating network objectives.

\subsection{Mobility, Handover, and Service Migration}
\label{subsec:mobility}

The LEO topology changes continuously, so a semantic session lasting more than a few minutes may experience a handover. Unlike conventional handover, semantic handover must also preserve the state required to encode and decode information. This state can include codec weights, codebook versions, knowledge-base revisions, and generative context. Current mobility procedures do not explicitly account for these semantic states, creating additional coordination requirements during handover and service migration. Two works address this problem from different perspectives. For vertical handover in an integrated sensing and communication SAGIN,~\cite{10886920} considers jitter, load imbalance, and robustness. It replaces the policy network in a multi-agent proximal policy optimization framework with a denoising diffusion model to generate handover decisions in a high-dimensional and time-varying action space. For service migration in satellite-aided UAV networks, Jia \emph{et al.}~\cite{11303273} address the limited global observability caused by rapidly changing topology. Their framework propagates local semantic features through cyclic recurrent message passing, allowing each agent to estimate network states that are not directly observable. Dueling deep Q-network agents then make distributed migration and routing decisions. Service migration can also be integrated directly into semantic transmission. In~\cite{huang2026semantic}, ISLs support both data forwarding and task migration. When the serving satellite lacks the computing resources needed to generate the requested quality, the generation task is transferred to another satellite with sufficient capacity.

\subsection{Delay- and Disruption-Tolerant Semantic Scheduling}
\label{subsec:dtn}

When propagation delays exceed the available control interval, network optimization shifts from selecting paths to making better use of each contact opportunity. The goal-oriented framework of Uysal~\cite{uysal2024goal} treats each contact window as a limited shared resource. Transmitting stale or task-irrelevant data not only wastes that opportunity but can also occupy relay buffers and delay information that is still useful. Semantic prioritization therefore becomes an end-to-end networking problem rather than only a transmission scheduling problem. The Internet of Space vision~\cite{IoSSemCom} extends this idea to missions with different semantic requirements, including real-time, archival, and security-critical applications. Rather than using a separate SemCom stack for each application, it proposes a common framework in which semantic encoders and decoders can be exchanged and composed according to application needs. Neither work implements a specific semantic codec. Instead, their main contribution is to make the delivered task value per contact an explicit networking objective, whereas the routing approaches in Section~\ref{subsec:routing} consider this objective only indirectly. The works summarized in Table~\ref{tab:networking} point to a common insight: an intermediate node in a semantic network is more than a repeater. It may forward, correct, re-encode, or translate the semantic representation, or move the computation to another node. What the intermediate node does with the representation is therefore a key design choice. The ``relay operation'' column captures this choice and highlights the main difference among the reviewed designs.

\begin{table*}[!t]
\centering
\caption{\textcolor{black}{Summary of Semantic Networking and Multi-Hop Transport Approaches for NTN. }}
\label{tab:networking}
\scriptsize
\renewcommand{\arraystretch}{1.10}
\setlength{\tabcolsep}{2.6pt}
\begin{tabular}{|>{\centering\arraybackslash}p{1.45cm}|>{\centering\arraybackslash}p{0.7cm}|c|c|>{\centering\arraybackslash}p{2.1cm}|>{\centering\arraybackslash}p{2.2cm}|>{\centering\arraybackslash}p{1.8cm}|p{7.5cm}|}
\hline
\textbf{Category} & \textbf{Ref.} & \textbf{Plat.} & \textbf{Par.} & \textbf{Relay operation} & \textbf{Path model} & \textbf{Reported metric} & \multicolumn{1}{c|}{\textbf{Mechanism and principal limitation}} \\ \hline

\multirow{2}{1.5cm}{\centering\textbf{Distortion accumulation}}
& \cite{10345598} & \xmark & DJ & Decode and re-encode & Generic multi-hop & Reconstruction quality & Recursive training on already-degraded receptions teaches the codec its downstream input distribution. Reduces but does not remove per-hop loss. \\ \cline{2-8}
& \cite{LinSemForward} & \xmark & DJ & Extract and forward semantics & Cooperative relay & Euclidean distance & Relay extracts and forwards semantics instead of full decode-and-forward. Relay must host the full codec. \\ \hline

\multirow{2}{1.5cm}{\centering\textbf{Latent-domain forwarding}}
& \cite{SemForwardCodebook} & S & DJ & Forward codebook indices & LEO satellite-terrestrial & PSNR, MS-SSIM & Forwarder manipulates codebook indices, not tensors; FiLM restores channel adaptivity; codebook supports model-division access. All nodes need the same codebook. \\ \cline{2-8}
& \cite{AerialSemRelay} & G & DJ & Forward or transcode per user & Satellite-UAV-multi-cluster & Sum rate, SE & Dual-mode relay forwards to capable users and transcodes for conventional ones. Transcoding cost at the UAV unmodelled. \\ \hline

\multirow{2}{1.5cm}{\centering\textbf{Gateway-assisted enhancement}}
& \cite{Nguyen2024NovelDenoising} & S & DJ & Latent-space denoising at gateway & Multi-gateway hop & Semantic similarity & Low-SNR users routed via a gateway that denoises in the latent domain. Gains concentrate at cell edge; adds a hop. \\ \cline{2-8}
& \cite{10901128} & H & DJ & Forward with allocated compute & Satellite-HAPS-ground & Quality of semantics & Relay computation, power, bandwidth, and forwarded content optimized jointly. HAPS treated as a generic relay, not a compute node. \\ \hline

\multirow{3}{1.5cm}{\centering\textbf{Semantic-aware routing}}
& \cite{GraphJSCR} & S & DJ & Optionally re-encode & Time-varying directed graph & Latency, drop rate & Factorized action selects hop, semantic budget, and relay re-encoding from local state. Trained per constellation; no transfer. \\ \cline{2-8}
& \cite{10740143} & S & KG & Encode only at AI satellites & Temporal graph, ILP & Acceptance ratio, throughput, delay & Feasible paths must cross an AI satellite whose KB matches the destination, which contact graph routing cannot express. KB matching binary. \\ \cline{2-8}
& \cite{10716597} & G & GAI & Agentic semantic signalling & Cross-domain SAGIN & Mission completion & LLM-driven nodes combine semantic signalling, routing, and intent-driven management. Visionary and magazine paper. \\ \hline

\multirow{3}{1.5cm}{\centering\textbf{Mobility \& migration}}
& \cite{10886920} & G & GAI & Handover decision & LEO-ground tiers & Jitter, load balance & Diffusion policy generates handover decisions in a high-dimensional action space; validated at 1000-2500 satellites. Semantic session state not migrated. \\ \cline{2-8}
& \cite{11303273} & G & KG & Migrate service, route request & Satellite-aided UAV graph & Delay, throughput & Recurrent message passing infers global state from local features for distributed DQN agents. Adds signalling over the same scarce links. \\ \cline{2-8}
& \cite{huang2026semantic} & S & GAI & Migrate the generation task over ISL & Direct satellite access & Semantic efficiency & ISLs treated as a task-migration plane: computation moves, not data. Meaningful only when payload is small relative to the model. \\ \hline

\multirow{2}{1.5cm}{\centering\textbf{Delay-tolerant scheduling}}
& \cite{uysal2024goal} & D & SIG & Contact-aware prioritization & Interplanetary, scheduled contacts & Information relevance & Contact-aware forwarding; stale data displaces useful data in relay queues. Framework-level, no codec. \\ \cline{2-8}
& \cite{IoSSemCom} & D & DJ & Cross-satellite orbits& Cross-mission Internet of Space & Qualitative & Argues heterogeneous missions need a common composable semantic substrate. Vision paper; no mechanism or evaluation. \\ \hline
\end{tabular}
\end{table*}

\section{Semantic Resource Management and Cross-Layer Optimization}
\label{sec:Resource}

Classical satellite resource allocation relies on relatively clear relationships between communication resources, such as bandwidth and power, and performance measures such as rate or reliability. In SemCom, this relationship is more difficult to characterize because the final performance also depends on the encoder and decoder, channel conditions, data, and downstream task. As a result, semantic or task quality at the receiver cannot generally be determined from bandwidth and power alone. SemCom also introduces the compression ratio as an additional decision variable. Changing the compression ratio affects not only the amount of data transmitted, but also the computation required for semantic extraction and reconstruction, as well as the resulting latency and energy consumption~\cite{ZhangSemRA,10437885}. Based on these characteristics, we organize the works around five main challenges: modeling the relationship between communication resources and semantic performance, balancing compression, computation, and energy, exploiting controllable network geometry, supporting coexistence with bit-level users, and scheduling transmissions according to information freshness.

\subsection{Recovering the Rate-Semantics Relation}
\label{subsec:ratesemantics}

Resource allocation requires a tractable relationship between the allocation decisions and the resulting semantic quality. Because this relationship is generally not available in closed form, existing approaches use different surrogate models to approximate it. We identify four strategies, which differ in their accuracy, data requirements, and ability to adapt to changing conditions.

\subsubsection{Empirical surrogate fitting}
The most direct approach is to measure semantic performance offline and fit a surrogate function to the results. For example,~\cite{nguyen2026joint} evaluates a trained codec across different compression ratios and SNR values and fits a function $\hat{Q}(\rho,\mathrm{SNR})$ to the measured performance. This function is then used in a conventional resource-allocation problem, allowing existing optimization methods to be applied after an offline characterization step. However, the fitted function is specific to the codec, dataset, and channel conditions used during evaluation. Changes in any of these factors, such as retraining the codec or a shift in the data distribution, may therefore reduce the accuracy of the surrogate.

\subsubsection{Analytic semantic metrics}

A second approach defines semantic metrics that can be computed directly from system performance. Examples include semantic transmission rate and spectral efficiency measured in semantic units~\cite{9763856}, as well as semantic energy efficiency~\cite{10460718}. NTN studies have also introduced composite metrics. For example, QoSem~\cite{10949600} combines semantic rate, accuracy, and processing time, while SemCom efficiency~\cite{11207608} combines reconstruction quality and delay. These metrics provide explicit objectives for optimization and a quantitative basis for comparing semantic and bit-level communication systems. However, composite metrics often depend on tunable weights, which can make comparisons less reliable. An apparent improvement may result from changing these weights rather than improving the underlying system performance. This issue is explicit in~\cite{huang2026semantic}, where the weights are themselves treated as optimization variables.

\subsubsection{Learned policies}
A third approach avoids explicitly modeling the relationship between resource allocation and task performance. Instead, it formulates resource allocation as a Markov decision process and learns a policy that maximizes long-term task performance. The choice of reinforcement learning algorithm mainly depends on the action space. Soft actor-critic (SAC) is used for continuous decisions such as power and bandwidth allocation~\cite{10901128,haarnoja2018soft}, while deep Q-networks (DQN) are used for discrete decisions such as user association and mode selection~\cite{XIE2025103762,wang2016dueling}. Proximal policy optimization (PPO) is commonly used for trajectory control~\cite{FedSemUAV,schulman2017proximal}, while hybrid-action methods handle continuous and discrete decisions within the same framework~\cite{10437643,11231113}. These learned policies can handle non-convex allocation problems that are difficult to solve analytically. However, their performance may degrade when the operating conditions differ from those encountered during training.

\subsubsection{Joint codec-policy training}

The fourth strategy jointly trains the semantic codec and the resource-allocation policy. Resource constraints are included during training, allowing the encoder to learn representations that match the conditions under which the allocation policy operates. This approach is less common but promising because it directly optimizes the codec and allocation policy together, without relying on an intermediate performance model. The work in~\cite{11207608} provides a clear NTN example. However, joint training also tightly couples the two components. As a result, the codec and allocator cannot be easily developed, validated, or updated independently. This can be particularly challenging when different organizations operate or maintain the transmitter and receiver.

None of the four strategies has emerged as a general solution. Instead, the reviewed works suggest that different strategies are suitable for different settings. Empirical fitting is commonly used when the codec is fixed, and the main objective is to optimize energy or latency. Learned policies are more suitable when the system state is high-dimensional and changes over time. Analytic semantic metrics provide explicit performance measures and are mainly used to formulate optimization objectives and compare different system designs.

\subsection{Compression, Computation, and Energy Coupling}
\label{subsec:coupling}

The key resource-allocation problem in NTNs is the trade-off among compression, computation, and transmission energy in Eq.~\eqref{eq:onboard_energy}. Increasing the compression level reduces the number of transmitted symbols and therefore the transmission energy, but requires more computation for encoding. When computation and transmission share the same limited energy budget, the compression level must balance these two costs rather than minimize either one independently~\cite{10999070}. Zhao \emph{et al.}~\cite{11006980} provide the most detailed analytical treatment of this trade-off. They consider a satellite-to-UAV-to-ground downlink in which the satellite and UAV perform probabilistic semantic compression using a shared probabilistic graph. The computation cost is modeled by a piecewise overhead function $O_k(\rho_k)$. As the compression level changes, the encoder may need to evaluate more complex conditional probabilities over the graph, causing the computation cost to vary with the compression decision rather than remain fixed. The resulting joint optimization problem is decomposed into six subproblems, with closed-form solutions derived for the computation capacity and UAV altitude at each iteration. These results provide direct insight into how different system variables affect the optimal resource allocation.

Several studies examine the same trade-off among compression, computation, and energy, but use different objectives and solution methods. Liu \emph{et al.}~\cite{11310140} apply probabilistic semantic compression before offloading along a user-to-UAV-to-LEO path and report energy savings of up to $78.4\%$ compared with local processing. The framework in~\cite{zhao2025agentic} further considers whether a task should be offloaded and minimizes system-wide energy under latency and quality constraints. For satellite-UAV systems,~\cite{10901382} minimizes energy while requiring transmission to finish within the available contact window. In contrast,~\cite{11365133} minimizes satellite-to-user latency under accuracy and power constraints while also accounting for user CPU frequency. Despite their different objectives and system settings, these studies capture the same basic trade-off between computation and communication costs. However, their results are difficult to compare directly because they use different system models and evaluation settings.

\subsection{Controllable Geometry: Trajectory, Deployment, and Association}
\label{subsec:geometry}

The UAV literature plays a distinct role in this plane because the UAV position is itself a decision variable. Changing the trajectory affects both the semantic transmission conditions, such as channel quality, and the energy required for flight. Many studies therefore jointly optimize the UAV trajectory and semantic transmission parameters, but differ in the variables and constraints they consider. When only communication resources are considered,~\cite{10437885} jointly optimizes the semantic symbol count, transmit power, and bandwidth under a semantic quality-of-service constraint. When mobility is included,~\cite{10419536} uses cooperating SAC agents to jointly select the trajectory, symbol count, and transmit power for a data-gathering mission. Similarly,~\cite{10507296} jointly optimizes the trajectory, semantic information selection, and user association to maximize the total similarity across users. For spectrum-limited systems,~\cite{XIE2025103762} uses NOMA to superpose semantic streams on the same channel and jointly optimizes the symbol count, trajectory, and transmit power to maximize the semantic sum rate. Finally,~\cite{XU2025102555} considers a UAV acting as a mobile base station with unknown user demands and jointly adapts its trajectory and compression ratio while satisfying the UAV's endurance constraint.

We identify three variations that differ from the standard joint trajectory and resource-allocation formulation. First, the command-and-control design in~\cite{SemUAVIoT} considers non-negligible control delay, whereas other trajectory-based studies typically assume delay-free control. Devices encode information into importance-ordered latent representations, allowing the receiver to reconstruct the information from only a subset of the transmitted symbols. Meanwhile, the base station controls the UAV through acceleration commands. Second, the post-disaster framework in~\cite{10843775} does not rely on a learned policy. Instead, it assigns collection, encoding, and forwarding tasks to different UAVs and uses Monte Carlo tree search to schedule these tasks. Third,~\cite{zheng2026low} considers UAV swarms, where UAVs first perform semantic aggregation within each cluster and then use collaborative beamforming to form a virtual antenna array for forwarding. An LLM tunes the evolutionary search parameters to account for changes in the decision space as the UAV clustering changes.

A smaller group uses optimization methods based on explicit system models rather than learned policies.~\cite{siyun2025fair} jointly optimizes the UAV trajectory, beamforming, compression rate, and number of active fluid-cantenna ports to maximize the minimum equivalent rate. This is the only work in this group that explicitly considers max-min fairness. Other studies mainly optimize sum or average performance, providing less insight into the performance of users with poor channel conditions. Freshness is another objective considered in this setting. The framework in~\cite{10943268} determines whether semantic extraction should be performed at the UAV or the ground user and minimizes semantic-aware AoI using a Lyapunov-based policy. In contrast,~\cite{10437643} minimizes mission completion time by balancing reconstruction quality and computation energy.

The same joint-optimization principle applies when the controllable element is a relay rather than the sensing source. In~\cite{11231113}, UAVs equipped with millimeter-wave radar and vision sensors collect real-time information about user density, blockage, and link quality. A SAC agent then uses this information to jointly optimize task offloading and collaborative uplink beamforming. The reported results achieve a $92\%$ offloading success rate and reduce delay by $38.9\%$ compared with greedy baselines.

\subsection{Multiple Access and SemCom-BitCom Coexistence}
\label{subsec:coexistence}
Semantic transmission is unlikely to fully replace bit-level transmission on operational satellites. The two will therefore need to share the available spectrum. This creates a new resource-allocation problem because semantic and bit-level users have different performance objectives that cannot be described by the same utility measure. Ahmed \emph{et al.}~\cite{Ahmed2025MultiSemCom} identify non-orthogonal multiple access as a natural solution. Through superposition, semantic and bit-level users can share the same resource block, while successive interference cancellation (SIC) separates their signals at the receiver. The power-allocation formulation in~\cite{11072816} follows this approach, but also reveals a key limitation: there is still no general model that directly relates semantic transceiver performance to the allocated communication resources. Rate-splitting multiple access (RSMA) may provide another option for semantic-bit-level coexistence. Its common stream could carry the most important semantic content, while private streams provide additional refinement. Other designs enable coexistence without superposition. The model-division approach in~\cite{SemForwardCodebook} separates users through their codebooks rather than power or frequency resources. The receiver distinguishes the streams according to the semantic model used for decoding, as different users maintain different semantic states. In contrast, the semi-OTFS scheme in~\cite{11423904} separates users according to their mobility classes. This distinction is important for SIC-based access because successful interference cancellation depends on the decoding order, whereas a semantic stream may not be directly decodable by a bit-level receiver.

\subsection{Freshness- and Significance-Driven Scheduling}
\label{subsec:freshness}

In status-update applications such as telemetry, infrastructure monitoring, and EO of transient events, the goal is not to deliver as much data as possible, but to keep the receiver updated with the most recent information. For example, a flood map is most useful while the flood is occurring. Age of Information~\cite{kaul2012}, version AoI, and AoMI~\cite{VisualEventLEO,EHAoI,SemAware6G} capture this requirement by measuring the freshness of the information available at the receiver. A freshness-aware scheduler therefore prioritizes recent updates over outdated ones. Semantic scheduling adds another criterion: an update should be transmitted when the underlying state has changed significantly, rather than simply because the previous update has become old. Delfani and Pappas~\cite{EHAoI} apply this idea to an energy-harvesting device reporting to an interconnected LEO network. By minimizing version AoI and avoiding transmissions when there is no new information, they report energy-efficiency improvements of up to $73\%$. The framework in~\cite{SemAware6G} further considers the trade-off among freshness, semantic relevance, and utility across network tiers, while~\cite{10943268} jointly optimizes semantic-aware AoI, UAV trajectory, and semantic extraction placement. These studies focus mainly on scheduling rather than semantic encoder design, and their objectives cannot be fully captured by conventional bit-based metrics. One limitation, however, is that the computation required to determine whether an update is semantically important is often excluded from the scheduling cost. Among the reviewed works,~\cite{11006980} and~\cite{EHAoI} explicitly account for this computation cost. This cost can become important for energy-harvesting systems and satellites with limited energy availability during eclipse periods.

\begin{table*}[!t]
\centering
\caption{\textcolor{black}{Summary of Semantic Resource Management Approaches for NTN.}}
\label{tab:resource}
\scriptsize
\renewcommand{\arraystretch}{1.1}
\setlength{\tabcolsep}{2.5pt}
\begin{tabular}{|>{\centering\arraybackslash}p{1.45cm}|>{\centering\arraybackslash}p{0.8cm}|c|c|c|>{\centering\arraybackslash}p{3.5cm}|>{\centering\arraybackslash}p{1.85cm}|p{7.5cm}|}
\hline
\textbf{Category} & \textbf{Ref.} & \textbf{Plat.} & \textbf{Par.} & \textbf{Surr.} & \textbf{Decision variables} & \textbf{Objective} & \multicolumn{1}{c|}{\textbf{Solution method and principal limitation}} \\ \hline

\multirow{4}{1.45cm}{\centering\textbf{Rate-semantics surrogate}}
& \cite{nguyen2026joint} & S & DJ & FIT & Compression ratio, power & Accuracy vs resource cost & Offline grid fit $\hat{Q}(\rho,\mathrm{SNR})$ feeding classical convex allocation. Invalid after any retraining or domain shift. \\ \cline{2-8}
& \cite{10949600} & H & DJ & MET & Power, bandwidth, compute, symbol selection & QoSem & Mixed-action DRL over a composite semantic metric. Composite weights are hand-set. \\ \cline{2-8}
& \cite{10437885} & U & DJ & MET & Semantic symbols, power, bandwidth & Semantic QoS vs resource use & Joint maximization of a QoS metric against resource consumption. Single-UAV, static users. \\ \cline{2-8}
& \cite{10901128} & H & DJ & RL & Compute, power, bandwidth, semantic content & Quality of semantics & Soft actor-critic over the satellite-HAPS-ground downlink. Policy tied to the trained topology. \\ \hline

\multirow{5}{1.45cm}{\centering\textbf{Compression -computation -energy}}
& \cite{11006980} & G & KG & MET & UAV location, beamwidth, bandwidth, power, $\rho_k$, task split & Total energy & Piecewise $O_k(\rho_k)$ makes computation an explicit function of compression; closed forms for capacity and altitude. Graph assumed current. \\ \cline{2-8}
& \cite{11310140} & G & KG & RL & Offloading, compression, resources, UAV deployment & Delay and energy & SCA deployment plus MAPPO offloading; 78.4\% energy reduction. Baseline is local processing, not a semantic competitor. \\ \cline{2-8}
& \cite{zhao2025agentic} & U & DJ & MET & Location, $\rho$, power, binary inference offload & System energy & Agentic controller over a mixed-integer non-convex problem. Agent overhead not accounted. \\ \cline{2-8}
& \cite{10901382} & G & KG & MET & UAV positions, $\rho$, compute-comm split & Energy under window constraint & Block-coordinate descent under coverage and visibility constraints. Window model simplified. \\ \cline{2-8}
& \cite{11365133} & S & DJ & FIT & Accuracy target, power, user CPU frequency & Latency & Latency minimization under an accuracy constraint. Single-hop, single-task. \\ \hline

\multirow{8}{1.45cm}{\centering\textbf{Controllable geometry}}
& \cite{10419536} & U & DJ & RL & Trajectory, symbols, power & Task accuracy, energy & Four cooperating SAC agents for single-UAV data gathering. Agent count fixed to the variable structure. \\ \cline{2-8}
& \cite{XIE2025103762} & U & DJ & RL & Symbols, trajectories, power, clustering & Semantic sum rate & NOMA superposition with $K$-means grouping and dueling DDQN. Sum-rate objective hides cell-edge users. \\ \cline{2-8}
& \cite{XU2025102555} & U & DJ & RL & Flight path, compression ratio & Mission completion & UAV serves users of unknown location and demand within endurance. Demand model stationary. \\ \cline{2-8}
& \cite{SemUAVIoT} & U & DJ & RL & Acceleration commands, symbol subset & Reconstruction quality & Importance-ordered latents allow partial reception; control downlink modelled with delay, unlike other trajectory work. Single UAV. \\ \cline{2-8}
& \cite{10943268} & U & SIG & RL & Extraction placement, trajectory & Semantic-aware AoI & Lyapunov-guided DRL with AO/SCA deployment refinement. Extraction energy at the ground user unmodelled. \\ \cline{2-8}
& \cite{zheng2026low} & U & DJ & MET & Clustering, positions, excitation weights, symbols/word & Rates and flight energy & Clustered UAVs form a virtual antenna array; an LLM tunes evolutionary search under clustering-dependent dimensionality. LLM cost unreported. \\ \cline{2-8}
& \cite{siyun2025fair} & U & DJ & MET & Trajectory, beamforming, $\rho$, antenna ports & Min equivalent rate & The only max-min fairness objective in this plane. Fluid-antenna hardware assumed available. \\ \cline{2-8}
& \cite{11231113} & G & \xmark & RL & Offloading, uplink beamforming & Success rate, delay & Perception-driven hybrid-action SAC with constraint masking; 92\% success, 38.9\% delay reduction. Perception overhead treated as free. \\ \hline

\multirow{3}{1.45cm}{\centering\textbf{Access \& coexistence}}
& \cite{11072816} & \xmark & DJ & MET & Power, beamforming, semantic symbol factor & Power efficiency & NOMA superposition of semantic and bit users; symbol factor by exhaustive search absent a transceiver model. Does not scale with users. \\ \cline{2-8}
& \cite{SemForwardCodebook} & S & DJ & \xmark & Codebook assignment & Multi-stream capacity & Model-division access separates users by which model decodes them. Requires distinct per-user models. \\ \cline{2-8}
& \cite{11423904} & G & DJ & MET & Mobility-class assignment & Rate region & Semi-OTFS access serves high- and low-mobility users without SIC. Two mobility classes only. \\ \hline

\multirow{3}{1.45cm}{\centering\textbf{Freshness scheduling}}
& \cite{EHAoI} & S & SIG & MET & Update decision & Version AoI, energy & Device updates only on version change; up to 73\% efficiency gain. Extraction energy only partly accounted. \\ \cline{2-8}
& \cite{SemAware6G} & G & SIG & MET & Sampling, scheduling & Freshness, relevance, utility & Generalizes the freshness-relevance-utility trade-off across TN-NTN tiers. Framework-level. \\ \cline{2-8}
& \cite{VisualEventLEO} & S & SIG & MET & Feature subset, transmission timing & AoMI & AoI analysis gives the fraction of users meeting timeliness for event classification. Event set fixed. \\ \hline
\end{tabular}
\end{table*}

In conclusion, the works in Table~\ref{tab:resource} address the same underlying problem, but use different approaches to connect resource allocation with semantic quality. The ``surrogate'' column summarizes this choice: FIT denotes empirical curve fitting, MET an analytic semantic metric, RL a learned policy, and JNT joint codec-policy training. Because there is no closed-form relationship between resource allocation and semantic quality, each study introduces a surrogate to capture this relationship. In practice, the choice of surrogate can have a greater influence on the system design than the optimization algorithm itself. Three observations emerge from the reviewed works. First, the coupling among compression, computation, and energy is widely recognized and depends strongly on the platform, but it has rarely been validated on real hardware. The optimal operating point therefore remains uncertain, which limits practical mission design. Second, UAVs are particularly suitable for joint trajectory and compression optimization because their geometry can be controlled. This flexibility partly explains why UAV-based studies dominate the literature. Third, most studies optimize sum or average performance, while only~\cite{siyun2025fair} considers a max-min objective. The reported gains therefore provide little information about users with poor channel conditions, even though serving such users is an important motivation for SemCom.

\section{Distributed Learning and Knowledge-Base Management}
\label{sec:Learning}

The semantic mechanisms reviewed so far rely on shared state between the transmitter and receiver. This state may include matched codec weights in D-JSCC, a shared codebook in digital semantic communication, a generative prior and its conditioning scheme in GAI SemCom, or an explicit knowledge base in reasoning-based SemCom. This section examines how such shared state is created, updated, and kept consistent across an NTN fleet, where nodes observe different data, communicate intermittently, and may be operated by different organizations. These conditions make distributed learning and state management more challenging than in terrestrial networks. Data collected by different nodes cannot always be centralized, while intermittent connectivity requires training and synchronization to be completed within available contact windows rather than through continuous connections. We therefore organize the literature into four areas: model training, task and domain heterogeneity, computation partitioning, and knowledge-base maintenance.

\subsection{Hierarchical Federated Learning over Orbital Topologies}
\label{subsec:hfl}

Federated learning (FL)~\cite{mcmahan2017} enables distributed nodes to train a shared model without sharing their local data. In NTNs, however, the aggregation hierarchy is determined by orbital topology and visibility rather than by implementation convenience. SemSpaceFL~\cite{SemSpaceFL} illustrates this structure. Satellites first train locally and upload their updates to a regional gateway during their visibility windows. Each gateway then aggregates the received updates, and a cloud server aggregates the gateway models. This hierarchy avoids requiring every satellite to communicate directly with a single server in the same training round. Instead, aggregation is performed at gateways that are accessible to groups of satellites with overlapping visibility. The same principle has been applied to other NTN tiers. NomaFedHAP~\cite{10438925} places parameter servers on HAPS, whose quasi-stationary positions provide more stable aggregation points than LEO satellites. Similarly,~\cite{FedNTNHFL,FLSTRA} organize aggregation according to the hierarchical structure of the NTN. Broader studies of FL in 6G NTNs~\cite{10478029,10679111} focus on a common scheduling problem: selecting the nodes that participate in each training round under time-varying connectivity and known visibility schedules.

\subsection{Task and Domain Heterogeneity}
\label{subsec:heterogeneity}

\begin{table*}[!t]
\centering
\caption{\textcolor{black}{Summary of Distributed Learning and Knowledge-Base Management Approaches for NTN.}}
\label{tab:learning}
\scriptsize
\renewcommand{\arraystretch}{1.10}
\setlength{\tabcolsep}{2.6pt}
\begin{tabular}{|>{\centering\arraybackslash}p{1.6cm}|>{\centering\arraybackslash}p{0.85cm}|c|c|>{\centering\arraybackslash}p{3cm}|>{\centering\arraybackslash}p{1.9cm}|p{7.5cm}|}
\hline
\textbf{Category} & \textbf{Ref.} & \textbf{Plat.} & \textbf{Par.} & \textbf{Aggregation structure} & \textbf{Reported metric} & \multicolumn{1}{c|}{\textbf{Mechanism and principal limitation}} \\ \hline

\multirow{4}{1.6cm}{\centering\textbf{Hierarchical FL over orbits}}
& \cite{SemSpaceFL} & S & DJ & Satellite $\to$ gateway $\to$ cloud & MS-SSIM, training time & Gateway-level aggregation exploits shared visibility among co-planar satellites; converges within a few orbits. Fixed gateway assignment. \\ \cline{2-7}
& \cite{10438925} & H & DJ & Devices $\to$ HAPS server & Convergence, overhead & HAPS parameter server with NOMA model exchange supplies the stable aggregation point no LEO node offers. Single-HAPS coverage limit. \\ \cline{2-7}
& \cite{FedNTNHFL} & H & DJ & HAPS servers with GEO/MEO relay & Convergence & Aggregation topology mirrors the NTN tier structure. Relay capacity treated as unconstrained. \\ \cline{2-7}
& \cite{10679111} & G & DJ & Adaptive across SAGIN tiers & Accuracy, handover cost & Orchestrates participation with adaptive offloading under changing visibility. Not specific to semantic models. \\ \hline

\multirow{3}{1.6cm}{\centering\textbf{Task \& domain heterogeneity}}
& \cite{TaskClusteringFL} & S & DJ & Cluster-wise by task similarity & Task accuracy & Similarity clustering with a transitivity constraint prevents destructive averaging across conflicting tasks. Threshold set manually. \\ \cline{2-7}
& \cite{nguyen2025cross} & \xmark & DJ & Domain-aware aggregation & Cross-domain accuracy & Representation alignment holds the shared latent space consistent while domain components aggregate separately. Domains assumed known. \\ \cline{2-7}
& \cite{11096603} & \xmark & DJ & Cross-modal federation & Accuracy & Extends federated semantic training across modalities. Not evaluated under intermittent connectivity. \\ \hline

\multirow{3}{1.6cm}{\centering\textbf{Split \& pruned training}}
& \cite{10445211} & S & DJ & Split satellite/ground with pruning & Accuracy, training cost & Pruning-split federated learning keeps raw data on board while improving delay, energy, and privacy. Split point fixed. \\ \cline{2-7}
& \cite{FedSemUAV} & U & DJ & Edge users train, UAV aggregates & Energy-delay product, fairness & Relieves the most energy-constrained node of training; trajectory and bandwidth optimized under fairness. Requires capable ground users. \\ \cline{2-7}
& \cite{XuUAVSwarm} & U & DJ & Intra-cluster centralized, inter-cluster decentralized & Convergence, robustness & Cluster heads combine centralized and decentralized federation; an evolutionary game sustains participation. Clustering overhead unreported. \\ \hline

\multirow{3}{1.6cm}{\centering\textbf{Knowledge-base maintenance}}
& \cite{LiuFangyu} & S & GAI & Point-to-point shared KB & Keyframe distance & Keyframe update resynchronizes the decoder reference, the only explicit maintenance mechanism reviewed. Update cost competes with payload. \\ \cline{2-7}
& \cite{11006980} & G & KG & Satellite/UAV $\to$ multi-GT broadcast & Energy efficiency & Compression derives from recovery against a shared probabilistic graph, so graph currency drives the reported savings. Aging not examined. \\ \cline{2-7}
& \cite{10740143} & S & KG & No aggregation mechanism & Acceptance ratio & Establishes KB compatibility as a hard routing feasibility constraint, not a quality parameter. Compatibility modeled as binary. \\ \hline
\end{tabular}
\end{table*}
Hierarchical aggregation is effective only when participating models can benefit from sharing information. In an NTN constellation, this assumption may not hold because satellites can differ in their learning tasks or data domains. The first case is task heterogeneity. When satellites perform different tasks, such as segmentation, classification, and detection, directly averaging their models can reduce performance across tasks. The federated cooperative multi-task framework in~\cite{TaskClusteringFL} addresses this problem through semantic-aware task clustering. It groups similar tasks based on a threshold rule and a transitivity constraint, ensuring that each task belongs to only one cluster. Model aggregation is then performed only among participants with compatible objectives. This approach is well suited to distributed satellite systems, where each satellite maintains its own encoder and cannot directly adopt a centralized multi-task formulation. 

The second case is domain heterogeneity, which can occur even when all satellites perform the same task. Satellites in different orbits may observe different environments, such as deserts, forests, oceans, and urban areas. A single model obtained by averaging models trained on these domains may perform worse than models adapted to individual domains. The cross-domain framework in~\cite{nguyen2025cross} addresses this problem through global representation alignment and domain-aware aggregation. It aligns the shared representation space across domains while aggregating domain-specific components separately. Related work has also extended federated training to cross-modal systems~\cite{11096603}.

\subsection{Split, Pruned, and Partitioned Training under SWaP}
\label{subsec:split}
FL keeps data local, but each participant must still train the model. For small satellites or UAVs, the required computation and energy may exceed the available resources. Existing work addresses this constraint in three main ways. The first approach moves training away from the resource-constrained platform. In~\cite{FedSemUAV}, edge users collaboratively train the semantic codec, while the UAV acts only as an aggregator. This reduces the computational and energy burden on the UAV. Its trajectory and bandwidth allocation are then jointly optimized to minimize the energy-delay product. The second approach partitions the model across nodes. The satellite-borne edge-cloud framework in~\cite{10445211} uses adaptive pruning-split federated learning to divide the network between the satellite and the ground. The on-board portion is pruned to fit the available resources. This reduces training delay and energy consumption while keeping the original data local, with only intermediate activations exchanged across the link. The third approach adapts the federation topology. For UAV swarms with highly dynamic links,~\cite{XuUAVSwarm} organizes UAVs into clusters, using centralized FL within each cluster and decentralized aggregation among cluster heads. This reduces coordination overhead within clusters while improving robustness to link changes across the swarm. Cluster selection is further formulated as an evolutionary game to encourage continued participation among self-interested UAVs. Overall, these approaches address three different challenges: limited on-board resources, large model size, and unstable network topology. In mobile NTN systems, however, participation may depend not only on connectivity and scheduling, but also on whether individual nodes have sufficient incentive to remain involved in the learning process.

\subsection{Knowledge-Base Construction, Synchronization, and Drift}
\label{subsec:kb}
The preceding subsections focus on model weights. A KB is a different system component: it is queried rather than optimized through training, and it must be updated as the underlying world or context changes. In reasoning-based and generative semantic communication, the KB can also provide much of the prior information that enables compression~\cite{10318078,ChristinaLessdata}. However, its construction and maintenance remain largely unexplored in the NTN literature. Probabilistic semantic compression schemes rely on a shared graph that allows the receiver to recover information omitted from transmission~\cite{11006980,11310140}. Their compression performance therefore depends on the KB remaining sufficiently current, yet neither work examines how performance changes as the graph becomes outdated. The audiovisual satellite system in~\cite{LiuFangyu} provides the only reviewed example of an explicit maintenance mechanism, updating keyframes to keep the decoder's reference aligned with the source. On the networking side,~\cite{10740143} treats KG compatibility as a requirement for path feasibility rather than simply a factor affecting communication quality, highlighting the operational importance of maintaining compatible knowledge bases. The main gap is the absence of a KB lifecycle. None of the reviewed studies addresses how a knowledge base should be initialized on a newly deployed node, monitored for drift during operation, or reconciled after different nodes diverge. Instead, most existing designs assume that the required knowledge is already available, correct, and consistent at both ends. Section~\ref{subsec:ch_kb} discusses the four operations required to address these issues, and Table~\ref{tab:learning} provides a summary of discussed works within the section.

\section{Security, Privacy, and Safety of Semantic NTN}
\label{sec:Security}

Security, privacy, and safety cut across all five planes rather than belonging to a single one. Different parts of the system introduce different risks: an eavesdropper may target the semantic representation, a compromised relay may disrupt network operation, and a poisoned update may corrupt the learning process. Other failures can occur without an external attacker; for example, a generative decoder may produce incorrect or misleading content. SemCom therefore inherits the vulnerabilities of the DL models it relies on, while also introducing risks specific to systems that transmit and reconstruct meaning rather than bits alone~\cite{SecureSemSurvey,SemanticSurvey0,DWonSurvey}.

\subsection{The Threat Surface}
\label{subsec:threats}

We consider three threats that apply to SemCom in general and are amplified by the NTN environment rather than created by it. Together, they threaten message confidentiality, inference integrity, and knowledge-base privacy.

\emph{Semantic eavesdropping} occurs when a passive attacker intercepts a semantic representation and attempts to recover the source or task-relevant information using a stolen or surrogate decoder. Unlike a conventional bit stream, a semantic representation is designed to retain information relevant to a particular task. An intercepted semantic stream may therefore provide the attacker with more directly usable information than an equivalent number of intercepted bits.

\emph{Adversarial perturbation} targets the learned components of the semantic system. Small, carefully designed changes to the input or transmitted symbols can cause the receiver to produce an incorrect task output~\cite{SecureSemSurvey}. Because these perturbations exploit the model rather than channel errors, conventional error-correction mechanisms may not detect them. The aggressive compression used in SemCom can further limit the redundancy available to identify or correct such manipulations.

\emph{Knowledge-base inversion and inference} targets the shared knowledge or learned models used by the system. An attacker with access to an updated knowledge base or model may attempt to reconstruct training data through model inversion or determine whether a particular sample was used during training through membership inference. This can be particularly sensitive when the training data contain imagery of sovereign territory. Techniques such as differential privacy and gradient encryption can reduce these risks.

\subsection{NTN-Specific Amplification}
\label{subsec:amplification}
NTNs can make existing security threats more difficult to detect and mitigate. The main reasons are the multi-hop communication path, the involvement of multiple operators, and the broadcast nature of many NTN links. Together, these properties create security challenges that differ from those in terrestrial networks.

The multi-hop path, for example, creates a risk of \emph{relay compromise}. A malicious intermediate node can modify the features it forwards and thereby alter the task output at the destination without revealing where the corruption occurred. This problem is particularly challenging for latent-domain forwarding, where intermediate nodes do not reconstruct or interpret the transmitted representation. As a result, there may be no point along the path where the semantic content can be inspected, making it difficult to identify the compromised hop. Accordingly,~\cite{Hu2025SAGIN} identifies practical defenses against compromised relays as an underdeveloped area.

The multi-operator structure creates a \emph{long-baseline trust problem}. Different organizations may operate different parts of the communication path while still needing to establish and maintain trust in a shared semantic state. Long propagation delays can also make interactive challenge-response authentication inefficient or impractical on some NTN links. This setting therefore calls for delay-tolerant approaches to trust establishment, such as threshold cryptography or pre-distributed semantic anchors, rather than relying only on interactive authentication protocols.

The broadcast nature of NTN links also creates \emph{jamming asymmetry}. A ground-based attacker may disrupt a user-facing downlink with relatively modest transmission power. SemCom can be particularly vulnerable because aggressive compression reduces the redundancy available to tolerate or recover from short periods of interference. The multi-UAV framework in~\cite{10925607} addresses this problem by jointly optimizing UAV trajectories, user associations, and channel selection under active jamming. Its multi-agent DRL scheme enables UAVs to coordinate their actions to maintain semantic accuracy while avoiding interference.

\subsection{Defence Mechanisms}
\label{subsec:defences}

We group the reviewed defense mechanisms into three families according to where they are applied: the learned model, the physical layer, or the transmitted and shared information.

\subsubsection{Hardening the learned model}
The first family improves robustness at the model level. Adversarial training exposes the codec to carefully designed perturbations during training, while joint training of the source and channel encoder and decoder helps the system tolerate both adversarial perturbations and channel noise~\cite{Hu2023RobustSemCom,Peng2023RDeepSC}. A different approach in~\cite{PairedARN} uses paired adversarial residual networks at the transmitter and receiver. The transmitter-side module generates adversarial examples, while the receiver-side module removes both the adversarial perturbation and channel noise. The two modules are jointly optimized to balance adversarial robustness, semantic error, and resistance to eavesdropping. In the knowledge-graph setting,~\cite{10872947} applies adversarial robustness to the graph construction process.

\subsubsection{Exploiting the physical layer}

The second family protects semantic transmission by exploiting wireless-channel properties. Mission-oriented secure semantic satellite communication in~\cite{MissionSecureSat} uses a weighted fractional Fourier transform, where the transform order acts as a shared key. A receiver without the correct order observes a scrambled signal, whereas the intended receiver can recover the transmitted information. This approach is well suited to NTN because it does not require an additional feedback exchange. Other methods use friendly jamming to interfere with known eavesdroppers~\cite{11011492,10870357}, artificial noise to suppress signals in selected directions~\cite{10622708,11082318}, or reciprocal channel measurements for key generation~\cite{7393435}. Codebook-domain frequency hopping in~\cite{SemForwardCodebook} addresses jamming by spreading transmitted codebook indices across frequencies while preserving their semantic content.

\subsubsection{Privacy-preserving transmission and trust bootstrapping}
The third family protects either the transmitted semantic information or the shared state used by the system. For HAPS-to-ground broadcasting, Nkurunziza and Umehara~\cite{nkurunziza2026attention} apply Gaussian differential privacy to the extracted features and then use a learned transformation that only the intended receiver can invert. The reported results preserve reconstruction quality for the legitimate receiver while substantially degrading the eavesdropper's output. For multi-operator settings, hierarchical federation~\cite{SemSpaceFL} and blockchain-anchored multi-vendor training~\cite{elmahallawy2025decentralized} provide mechanisms for multiple organizations to contribute to shared semantic models or state without directly exposing their local data or models.

\subsection{Safety of Generative Decoders}
\label{subsec:safety}

Generative decoders introduce a safety risk that is distinct from adversarial attacks. Because they are optimized for plausibility, they can generate convincing but incorrect content when the received representation is severely corrupted. The decoder may provide no indication that the reconstructed content differs from the original source. For example, a diffusion decoder may generate a structure that was absent from an EO image. The resulting output could still appear realistic at the gateway and affect decisions based on that observation. Several safeguards could help address this problem. These include uncertainty estimation for decoder outputs, consistency checks against knowledge-graph constraints, and fallback to conventional bit-level transmission when decoder confidence is low. Although these techniques have each been studied individually, the reviewed literature has not combined or evaluated them for generative semantic communication in an NTN setting. A related risk is prompt injection through untrusted content~\cite{GreshakeLLM}. This is particularly relevant to LLM-agent designs, where the agent's instructions and the information it processes may traverse the same communication channel.

\section{Standardization, Testbeds, and Open Resources}
\label{sec:Standards}

For NTN-SemCom to move beyond research prototypes, its mechanisms need to be compatible with standardized interfaces, evaluated under agreed channel models, and supported by reproducible implementations or datasets. This section examines the current status of NTN-SemCom along these three dimensions.

\subsection{The 3GPP NTN Roadmap and SemCom Positions}
\label{subsec:3gpp}

3GPP introduced NTN access in Release~17, defining the architecture for transparent-payload satellite access to New Radio and specifying the associated propagation and channel models in TR 38.811~\cite{3gpp2020study,3gpp_ts23501_rel17,HosseinianSatelliteReview}. Release~18 further refined timing relationships, hybrid automatic repeat request operation, and mobility procedures to address long round-trip delays and rapidly changing cell coverage~\cite{3gpp_ts38300_rel18}. Release~20 is expected to introduce AI/ML-native features and deeper integration with sensing~\cite{11186253}. This creates a potential entry point for SemCom, although semantic-specific support is not currently defined. Three extensions would be particularly important. The first is a \emph{quality-of-service descriptor for semantic traffic}. Current 5G QoS flows are specified using parameters such as data rate, latency, and packet error rate. These parameters cannot directly express requirements such as maintaining task accuracy above a given threshold or semantic similarity above a target level. Consequently, the network cannot currently request or guarantee a semantic service level through a dedicated semantic bearer. The second is a \emph{coexistence rule} that defines how semantic and bit-level traffic share time-frequency resources. The multiple-access studies in Section~\ref{subsec:coexistence} show that superposition and model-division schemes can support such coexistence. However, scheduling these schemes requires the network to know which flows are semantic and what level of distortion their decoders can tolerate, and current signaling does not provide this information. The third is a \emph{model and knowledge-base identity mechanism}. The routing results of~\cite{10740143} show that semantic reachability depends on whether network nodes have compatible semantic state. Standardization would therefore need mechanisms to identify codec versions, codebook revisions, and knowledge-base states, and to negotiate their compatibility when a session is established. 
\begin{table*}[!t]
\centering
\caption{\textcolor{black}{Resources Used in the Reviewed NTN-SemCom Literature and Their Availability.}}
\label{tab:resources}
\footnotesize
\renewcommand{\arraystretch}{1.10}
\setlength{\tabcolsep}{4pt}
\begin{tabular}{|p{1.5cm}|p{5.3cm}|p{3.8cm}|p{6.5cm}|}
\hline
\textbf{Resource category} & \textbf{Resources used in the reviewed works} & \textbf{Availability} & \textbf{Requirements for comparability} \\ \hline

Source data 
& Public Earth-observation and remote-sensing imagery; aerial and UAV image datasets; standard vision benchmarks used as proxies; text corpora used as telemetry proxies 
& Mostly public, but scene selection, spatial resolution, and preprocessing vary across studies 
& A common NTN-SemCom evaluation dataset with fixed scenes, task labels, and train/test splits, covering representative terrain and cloud conditions encountered by NTN systems 
\\ \hline

Channel models 
& AWGN; Rayleigh and shadowed-Rician fading; Fisher-Snedecor $F$~\cite{9959884}; M\'alaga turbulence for optical links~\cite{10969992}; 3GPP NTN-TDL profiles in one study~\cite{11423904} 
& Model specifications are publicly available through 3GPP TR 38.811 and ITU-R~\cite{3gpp2020study,series2015propagation} 
& Mandatory evaluation using standardized NTN channel profiles, with elevation angle, atmospheric attenuation, and Doppler variation over a complete satellite pass 
\\ \hline

Orbital and constellation models 
& Published Starlink and OneWeb constellation parameters~\cite{StarlinkConstellation,FCC_Starlink_2019}; idealized circular orbits; constellation propagators with different levels of fidelity~\cite{esswein2026constellation,10886920} 
& Public constellation parameters are available, but propagator implementations are rarely released 
& A reference pass generator that produces elevation, range, Doppler, and visibility traces from published ephemerides 
\\ \hline

On-board compute models 
& Analytical energy-compression models; processor comparisons from~\cite{lovelly2017comparative}; space-qualified AI accelerators surveyed in~\cite{ortiz2023onboard} 
& Processor and accelerator data are reported; measured semantic-encoder energy consumption on hardware is unavailable 
& Measured encoder energy consumption and latency on representative space-qualified accelerators to determine the practical optimum in~\eqref{eq:onboard_energy} 
\\ \hline

Metrics 
& PSNR, MS-SSIM, LPIPS, task accuracy, BLEU, semantic-unit metrics, AoI variants, and study-specific composite metrics such as QoSem and SCE 
& Metric definitions are published, but composite metric weights vary across studies 
& A minimum reporting standard comprising at least one fidelity or task metric, one cost metric, and the corresponding channel and orbital conditions, so that the trade-offs of each design can be assessed 
\\ \hline

Code and models 
& Released inconsistently; most reviewed studies do not provide trained codec weights or complete simulation configurations 
& Limited availability of source code, trained models, and evaluation configurations 
& Public release of trained codecs, model configurations, and evaluation settings to enable reproduction and reuse of the surrogate models discussed in Section~\ref{subsec:ratesemantics} 
\\ \hline

\end{tabular}
\end{table*}

\subsection{ITU-R, DVB, and Delay-Tolerant Networking Anchors}
\label{subsec:otherstandards}

Beyond 3GPP, three standards and frameworks define important constraints for NTN-SemCom. ITU-R Recommendation P.618 and related attenuation recommendations~\cite{series2015propagation,series2019attenuation} provide the propagation models used to estimate link availability and rain attenuation. They should therefore serve as reference models for evaluating SemCom under rain-induced fading. However, the reviewed adaptation studies mainly use these models to select channel parameters rather than to evaluate link availability using the corresponding ITU-R criteria. DVB-S2X~\cite{etsi_dvbs2x_2024} defines the framing, channel coding, and modulation used in many operational satellite systems. It therefore provides an important compatibility constraint for the digital semantic representations discussed in Section~\ref{subsec:discrete}: a semantic representation that cannot be carried over existing DVB-S2X infrastructure would require changes to the deployed ground segment. For deep-space communication, the Bundle Protocol and Contact Graph Routing provide the store-and-forward networking framework on which goal-oriented scheduling such as~\cite{uysal2024goal} can operate. Semantic prioritization in this setting would therefore need to be integrated with these existing mechanisms for managing scheduled and intermittent contacts.

\subsection{Datasets, Simulation Platforms, and Testbeds}
\label{subsec:resources}
Reproducibility remains a major challenge in NTNs-SemCom because existing studies use different datasets, channel models, and simulation settings, as shown in Table~\ref{tab:resources}. The main gap is the lack of a common evaluation setting. Remote-sensing imagery is the dominant source modality, but studies use different scenes, spatial resolutions, and preprocessing procedures. Channel models also vary widely, ranging from additive white Gaussian noise to Rayleigh fading, shadowed-Rician fading, and, in one case, standardized NTN-TDL profiles~\cite{11423904}. These differences make direct comparisons across studies difficult. Meaningful evaluation therefore requires a common benchmark with shared datasets, channel models, metrics, and computational assumptions. As Table~\ref{tab:resources} shows, the main resources needed to build such a benchmark are already publicly available.

\section{Open Challenges and Future Research Directions}
\label{sec:Challenges}

The literature reviewed in this survey shows that SemCom can address several constraints that are particularly important in NTNs, leading to mechanisms such as predictive rate control, latent-domain relaying, and orbit-scheduled federated training. However, significant challenges remain before these approaches can support reliable deployment. These challenges are not limited to algorithm design; they also involve verification, interoperability, resource constraints, and system integration. We identify five research challenges, each motivated by the findings discussed in the preceding sections.

\subsection{Challenge 1: Trustworthy Semantic Reconstruction}
\label{subsec:ch_verification}

Two seemingly different problems lead to the same issue: the receiver may obtain content that looks plausible but does not faithfully reflect the original observation. A generative decoder can produce such content when the received information is severely corrupted. An attacker can also manipulate latent features to produce a similarly plausible but incorrect reconstruction. This is particularly problematic for EO applications, where a synthesized feature may be difficult to distinguish from a real observation at the gateway. Downstream users in disaster response, scientific analysis, or insurance assessment may therefore have no reliable way to determine whether the received content reflects the original observation or a plausible reconstruction.

Addressing this problem requires three capabilities that are largely missing from the current literature. First, the reconstruction should be accompanied by a \emph{confidence signal}. The decoder should provide calibrated, region-level uncertainty to indicate which parts of the reconstruction are reliable and which are generated with low confidence. Among the reviewed works, only~\cite{ahmadi2025semantic}, in a different context, reports calibrated uncertainty. Second, the system needs a \emph{low-cost verification mechanism}. FMSAT's split coarse-fine error detection~\cite{FMSat} provides the closest example, but it detects transmission errors rather than incorrect content generated by the decoder. One possible approach is to transmit a compact verification digest with the semantic payload, such as low-order statistics or symbolic constraints that the reconstruction should satisfy. Third, a \emph{unified threat model} is needed. Existing defenses usually consider one attack type on one link. In practice, a multi-vendor constellation may simultaneously face jamming, compromised relays, and untrusted federation participants, while long propagation delays limit interactive authentication. This also makes security mechanisms that require additional round trips difficult to use in the high-delay scenarios where SemCom can be most useful.
\subsection{Challenge 2: Pass-Scale Semantic Adaptation}
\label{subsec:ch_ephemeris}

The predictive methods reviewed in Section~\ref{sec:ChannelAdaptive} mainly estimate future channel conditions from recent observations. However, part of the channel variation over a satellite pass can be predicted in advance from orbital geometry. Elevation, range, free-space path loss, and Doppler can be estimated from two-line elements before the pass begins, yet none of the reviewed works uses this information to plan transmission over the entire pass. This creates an opportunity to move beyond short-term rate adaptation toward \emph{pass-scale content scheduling}. Given a predicted pass trajectory and a queue of payloads with different task values, the transmitter could schedule high-value content during favorable channel periods and defer lower-priority payloads to weaker parts of the pass. The compression ratio of each payload could also be selected according to its predicted SNR. This problem is primarily one of scheduling and optimization rather than learning. The framework in~\cite{luo2026learning}, the only reviewed work that optimizes over the visibility window and includes the on-board queue in its state, provides a natural starting point. A direct extension could jointly optimize content ordering and compression ratio over the predicted pass. The same framework could also schedule low-priority knowledge-base synchronization during poor channel conditions, reserving the stronger parts of the pass for time-sensitive or high-value data.

\subsection{Challenge 3: Deployable Semantic Processing}
\label{subsec:ch_hardware}

Sections~\ref{sec:Representation} and~\ref{sec:Resource} reveal the same deployment gap from different perspectives. Although many studies acknowledge SWaP-C constraints, few evaluate semantic processing on hardware that represents actual flight platforms. In particular, the trade-offs among compression, computation, energy, and latency are usually derived analytically rather than measured on space-qualified hardware. Such models can indicate that an optimum exists, but they cannot determine where it lies on a real processor. The immediate need is therefore hardware measurement rather than another optimization algorithm. Representative semantic encoders, such as a sparse linear projector~\cite{CompressedLearning}, a quantized convolutional encoder, a distilled transformer, and a codebook lookup, should be evaluated for latency and energy consumption on radiation-hardened processors and space-qualified accelerators~\cite{ortiz2023onboard,lovelly2017comparative}. These measurements would provide the hardware parameters required by the formulations in Section~\ref{subsec:coupling} and help determine which semantic paradigms are feasible at different NTN tiers. They could also clarify where different processing methods should be deployed. For example, computationally intensive generative decoding may be better suited to ground infrastructure, whereas sparse or codebook-based methods may be more practical for on-board processing.

\subsection{Challenge 4: Semantic State \& Knowledge-Base Lifecycle}
\label{subsec:ch_kb}

The reviewed literature leaves a major gap in how semantic state is maintained over the lifetime of an NTN system. Representation schemes beyond continuous latents depend on receiver-side state, but existing studies generally assume that this state remains synchronized. Mobility studies consider handover decisions without specifying how semantic state is transferred or updated, while only one reviewed work explicitly addresses knowledge-base maintenance. Four operations are needed to make semantic state manageable in practice. \emph{Bootstrapping} defines how a newly launched satellite obtains the required codec, codebook, or knowledge-base state, potentially through pre-loading before launch followed by incremental synchronization during commissioning. \emph{Versioning} provides identifiers for codec, codebook, and knowledge-base revisions so that incompatible states can be detected when a session starts rather than causing silent quality degradation. \emph{Drift detection} identifies when the transmitter and receiver have diverged during operation. This is challenging because the resulting degradation can be difficult to distinguish from channel-induced errors. \emph{Reconciliation} determines how the two sides are brought back into agreement once divergence is detected, particularly when a full synchronization may not fit within a single contact window. Extending the binary compatibility model of~\cite{10740143} to a graded and version-aware compatibility model is therefore a concrete first step.

\subsection{Challenge 5: Native NTN Semantic Networking and Intelligence}
\label{subsec:ch_standards}

The remaining challenges concern how SemCom can be integrated into the network itself rather than treated as an additional application-layer capability. They include one standards challenge and two under-explored research opportunities. The main standards challenge is the lack of three basic mechanisms needed to support native semantic communication. First, a \emph{semantic QoS descriptor} is needed to define the quality level that a network can use for admission and resource allocation. Such a descriptor must be general enough to be signalled across different tasks while remaining relevant to the task being performed; the paper-specific composite metrics discussed in Section~\ref{subsec:ratesemantics} do not yet provide such a common definition. Second, a \emph{coexistence rule} is needed to allow a terminal to indicate the type and quality of semantic information its decoder can support. This would allow the network to schedule semantic and bit-level traffic together instead of treating semantic traffic as an opaque data stream. Section~\ref{subsec:coexistence} discusses mechanisms for such coexistence, but does not define the required signalling procedure. Third, a \emph{model and knowledge-base identity mechanism} is needed to identify the semantic models and knowledge bases used by communicating nodes. This requirement follows from the versioning and synchronization issues discussed in Section~\ref{subsec:ch_kb}; a standard identity mechanism cannot be defined independently of the underlying versioning scheme. Release~20's AI/ML-native scope provides a natural opportunity to address these issues, and we argue that these definitions should be established before the release is finalized.

\subsection{Future Directions}
The first opportunity is HAPS. Only a small number of studies consider HAPS, and most of them use it as a relay~\cite{10949600,10901128,nkurunziza2026attention}, even though Table~\ref{tab:platformvectors} shows that HAPS has the most favorable constraint vector among the considered platforms. HAPS could instead serve as a computing and storage layer for the NTN, hosting knowledge bases, generative decoders, or federated aggregation between the LEO and user tiers. Its position and available resources also make it a potential location for computationally intensive decoding that would be difficult to perform on user devices or satellites. This role has received little attention in the reviewed literature, suggesting that current studies have focused more on tractable relay-based formulations than on the broader potential of HAPS.

The second opportunity concerns foundation models. Existing studies use them mainly as semantic decoders~\cite{FMSat,LiuFangyu,10960413}, while their roles in networking and distributed learning remain largely unexplored. Three questions are particularly important. First, can a single foundation model serve as a receiver for heterogeneous transmitters, reducing the need for each transmitter to maintain a matched decoder and potentially simplifying the state-alignment problem discussed in Section~\ref{subsec:ch_kb}? Second, can natural language provide an efficient representation for the structured semantic information exchanged by reasoning-based SemCom~\cite{ChristinaLessdata}? Third, when is it more efficient to keep a foundation model at a gateway rather than distill it onto a HAPS or satellite?

These directions support the broader Internet of Space vision~\cite{IoSSemCom}, in which EO, navigation, direct-to-handset, deep-space, and on-orbit servicing applications share a common communication and computing infrastructure. In such a system, semantic encoders and decoders could be reused and updated across missions instead of being developed separately for each application. Achieving this vision requires several common foundations identified throughout this section: versioned shared state, agreed semantic quality measures, and standardized interfaces. These requirements are therefore as important as further advances in semantic codec design.

\section{Conclusion}
\label{sec:Conclusion}

This survey organizes semantic communication research in non-terrestrial networks by the design plane where each mechanism operates. We first identify five structural NTN constraints and examine how different semantic mechanisms address them, along with their associated costs. We then define platform-specific constraint vectors. These show that the relative severity of the constraints determines which mechanisms are suitable, and whether their benefits are worth the cost. Building on this analysis, we propose a five-plane taxonomy with a cross-cutting trust plane, and review the literature within this framework. Finally, we summarize the current standardization status and available resources, and identify five open challenges drawn from the gaps discussed in the preceding sections. Three conclusions emerge from this survey. First, the relationship between semantic communication and non-terrestrial networks is structural, not incidental. NTN constraints determine which semantic mechanisms are useful. They have also motivated mechanisms with no clear terrestrial counterpart, such as ephemeris-aware adaptation, latent-domain relaying, and orbit-scheduled federated learning. Second, the main limitations of the field are no longer purely algorithmic. Of the five open challenges identified here, one requires better measurement, one requires coordination, two require supporting infrastructure, and one requires standardization. Third, research effort has concentrated on problems with established optimization tools, at the expense of infrastructure that semantic communication actually needs. A large body of the reviewed works address resource allocation. Knowledge-base management, which underpins semantic consistency across several design planes, has received very little attention. As 3GPP moves toward AI/ML-native network operation and satellite constellations continue to grow, semantic communication could develop from a collection of specialized codecs into a broader framework for NTN networking. Priorities include energy-efficient semantic processing on space and aerial platforms, mechanisms for versioning and synchronizing shared semantic state, common evaluation conditions, and standardized interfaces for semantic information.

\bibliographystyle{IEEEtran}
\bibliography{mybib}

\end{document}